\documentclass[5p,authoryear]{elsarticle}

\usepackage[T1]{fontenc}
\usepackage[utf8]{inputenc}
\usepackage{lmodern}
\usepackage{amsmath,amssymb}
\usepackage{graphicx}
\usepackage{booktabs}
\usepackage{longtable}
\usepackage{array}
\usepackage{pdflscape}
\usepackage{xcolor}
\usepackage{microtype}
\usepackage[hyphens]{url}
\usepackage[hidelinks]{hyperref}

\graphicspath{{figures/}}
\newcommand{\NCpool}{190}
\newcommand{\NCtop}{100}
\newcommand{\NCdead}{32}
\newcommand{\NCliving}{68}
\newcommand{\NCflag}{19}
\newcommand{\NCwomen}{2}
\newcommand{\NCadded}{62}

\newcommand{\NCtied}{40}
\newcommand{\Ccutoff}{62.0}
\newcommand{\Cmedian}{69.0}
\newcommand{\Cmax}{93.0}
\newcommand{\NCmcsecure}{81}

\newcommand{\NCscsecure}{58}

\newcommand{\NCjtsecure}{54}

\newcommand{\NCtopten}{10}
\newcommand{\NCmc}{20{,}000}

\newcommand{\NCregNA}{61}
\newcommand{\NCregEU}{32}
\newcommand{\NCregAS}{3}
\newcommand{\NCregOC}{4}

\newcommand{\CbandLo}{0.30}
\newcommand{\CbandHi}{0.82}
\newcommand{\CsbandLo}{0.47}
\newcommand{\CsbandHi}{0.59}

\newcommand{\CovlImp}{96}

\newcommand{\CovlCit}{93}

\newcommand{\CovlNoR}{98}

\newcommand{\NCscreenP}{6}
\newcommand{\NCscreenA}{12}
\newcommand{\NCscreenB}{10}
\newcommand{\NCscreened}{28}
\newcommand{\NCidentified}{218}
\newcommand{\NRpool}{167}
\newcommand{\NRtop}{102}
\newcommand{\NRdead}{11}
\newcommand{\NRliving}{91}
\newcommand{\NRflag}{10}
\newcommand{\NRwomen}{12}
\newcommand{\NRadded}{35}

\newcommand{\NRtied}{23}
\newcommand{\Rcutoff}{64.0}
\newcommand{\Rmedian}{70.8}
\newcommand{\Rmax}{90.5}
\newcommand{\NRmcsecure}{87}

\newcommand{\NRscsecure}{60}

\newcommand{\NRjtsecure}{55}

\newcommand{\NRtopten}{8}

\newcommand{\NRregNA}{62}
\newcommand{\NRregEU}{25}
\newcommand{\NRregAS}{13}
\newcommand{\NRregOC}{2}

\newcommand{\RbandLo}{0.39}
\newcommand{\RbandHi}{0.57}
\newcommand{\RsbandLo}{0.51}
\newcommand{\RsbandHi}{0.57}

\newcommand{\RovlImp}{98}

\newcommand{\RovlCit}{93}

\newcommand{\RovlNoR}{99}

\newcommand{\NRscreenA}{10}
\newcommand{\NRscreenB}{5}
\newcommand{\NRscreened}{15}
\newcommand{\NRidentified}{182}
\newcommand{\CrhoEraS}{0.22}
\newcommand{\RrhoEraS}{0.26}
\newcommand{\Ncross}{10}
\newcommand{\Nentries}{192}
\newcommand{\Nunique}{193}
\newcommand{\NCnatTh}{83}

\newcommand{\NCnatHw}{13}

\newcommand{\NCnatEd}{32}
\newcommand{\CeraPeak}{1980}
\newcommand{\NCeraPeak}{29}
\newcommand{\NRnatTh}{41}

\newcommand{\NRnatHw}{59}

\newcommand{\NRnatEd}{28}
\newcommand{\ReraPeak}{2000}
\newcommand{\NReraPeak}{29}
\newcommand{\rhoMin}{0.94}
\newcommand{\rhoMax}{0.99}
\newcommand{\tauMin}{0.81}
\newcommand{\tauMax}{0.95}
\newcommand{\Npooltotal}{357}

\newcommand{\Nadded}{97}
\newcommand{\Nreassess}{80}
\newcommand{\Nsc}{20{,}000}

\newcommand{\NCSA}{45}

\newcommand{\NCSAin}{36}

\newcommand{\NBEL}{44}

\newcommand{\NBELin}{28}

\newcommand{\NQUA}{15}

\newcommand{\NQUAin}{12}

\newcommand{\NBOD}{37}

\newcommand{\NBODin}{26}

\newcommand{\NRAA}{23}

\newcommand{\NRAAin}{21}

\newcommand{\NPIO}{42}

\newcommand{\NPIOin}{36}

\newcommand{\Nrosterin}{159}
\newcommand{\Nrosterscored}{206}

\makeatletter
\newif\ifwidemode
\newcommand{\startwide}{\if@twocolumn\global\widemodetrue\onecolumn\else\global\widemodefalse\fi}
\newcommand{\stopwide}{\ifwidemode\twocolumn\global\widemodefalse\fi}
\makeatother

\journal{Annual Reviews in Control}

\begin{document}

\begin{frontmatter}

\title{The people behind feedback and machines: An impact-weighted review of the top~100 scholars in control systems and in robotics, 1926-2026}

\author[a,b]{Ali Nasir\corref{cor1}}
\ead{ali.nasir@kfupm.edu.sa}
\cortext[cor1]{Corresponding author.}
\affiliation[a]{organization={Control and Instrumentation Engineering Department, King Fahd University of Petroleum and Minerals},
            addressline={},
            city={Dhahran},
            postcode={31261},
            country={Saudi Arabia}}
\affiliation[b]{organization={Interdisciplinary Research Center for Intelligent Manufacturing and Robotics, King Fahd University of Petroleum and Minerals},
            addressline={},
            city={Dhahran},
            postcode={31261},
            country={Saudi Arabia}}

\begin{abstract}
Reviews in control and robotics usually organise a field by its ideas. This review organises the two fields by the people who produced those ideas, and uses that lens to synthesise how the ideas evolved, how they crossed between the two communities, and where research is heading. Candidates were identified from six award rosters and from histories, handbooks, roadmaps and obituaries; screened by documented rules (a faculty appointment, a principal contribution made between 1926 and 2026, and field scope), with every exclusion logged; and scored on six anchored criteria: impact on humanity (35\%), field creation (20\%), era-normalised scholarly influence (15\%), and technology transfer, education and peer recognition (10\% each). A source-backed scoring record accompanies all \Npooltotal{} scored candidates. The resulting lists name the top 100 scholars in control systems and in robotics; a three-way tie at the robotics cut-off gives \NRtop{} robotics entries, and with \Ncross{} bridge scholars the lists contain \Nentries{} entries representing \Nunique{} individuals. The rankings are stable under alternative weightings (Spearman $\rho$ between \rhoMin{} and \rhoMax{}), and inclusion frequencies under perturbed weights and scores identify which places near the cut-off are fragile. Agreement with award rosters is reported as a consistency check, not as validation. The synthesis traces the main eras of both fields, documents sixteen illustrative concept-migration pathways, and sets out a research agenda classified as most impactful, low-hanging fruit, natural progressions and wild cards. The scores are structured judgements; evidence, data and code are released for re-weighting and independent reassessment.
\end{abstract}

\begin{highlights}
\item Impact-first, source-backed scoring of 357 scholars in control and robotics
\item Rule-based screening; every exclusion is logged with a code and a source
\item Top-100 lists: 192 entries, 193 individuals, ten bridge scholars
\item Sixteen dated pathways show how concepts crossed between the two fields
\item Agenda: certified learning-enabled autonomy, grid control, assistive robots
\end{highlights}

\begin{graphicalabstract}
\includegraphics[width=\textwidth]{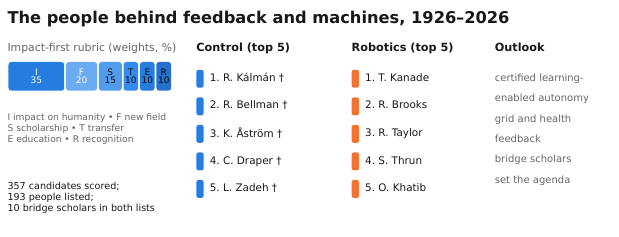}
\end{graphicalabstract}

\begin{keyword}
History of control \sep History of robotics \sep Research impact \sep Composite indicators \sep Scientometrics \sep Optimal control \sep Robot learning
\end{keyword}

\end{frontmatter}

\tableofcontents

\section{Introduction}\label{sec:intro}

\subsection{Background}

Feedback control and robotics are among the most consequential technologies of the past century. Estimation and guidance algorithms steered the Apollo spacecraft and now run in satellite-navigation receivers worldwide \citep{mcgee1985discovery,grewal2010applications}. Proportional-integral-derivative (PID) controllers make up the overwhelming majority of regulatory loops in the process industries, and relay auto-tuning developed in academia helped to automate their commissioning \citep{desborough2002increasing,astrom1984automatic,astrom2001future}. Model predictive control (MPC) runs refineries and chemical plants \citep{qin2003survey}, and the operational stock of industrial robots reached a record five million units in 2025 \citep{ifr2026world}. Surgical, rehabilitation and assistive robots, legged machines, aerial robots and warehouse fleets have followed.

The literature reviews these technologies by topic. Recent articles in this journal illustrate the genre: tutorials and surveys on nonsmooth robot-object systems \citep{brogliato2023modeling}, control barrier functions \citep{garg2024advances}, safe learning for multi-robot systems \citep{garg2024learning}, bio-inspired control of wearable robots \citep{almeida2025bio} and event-triggered control of autonomous robots \citep{yu2026bibliographical}, alongside method-oriented studies such as sampling-based path planning in cluttered environments \citep{venu2026optimized}. Each asks what is known about a class of methods. None asks who made it known, by what route ideas reached practice, and with what consequences for people.

This review asks those questions. It is organised around the scholars who created control systems and robotics as academic disciplines between 1926 and 2026, ranks them with an explicit rubric that gives the largest weight to impact on humanity, and uses the resulting lists to study how ideas evolved, where the two fields met, and where they are going.

\subsection{Why review people?}\label{sec:whypeople}

There are four reasons to review a field through its people. First, ideas travel with people: students carry methods between laboratories, and many concepts in this review crossed from control to robotics, or the reverse, because the same individuals worked on both sides (Section~\ref{sec:crossroads}). Second, the canon of both fields is written in names: the Kalman filter, the Bellman equation, Lyapunov functions, Denavit-Hartenberg parameters, the Lucas-Kanade tracker. Accurate attribution is part of teaching the canon, and it is fragile; our verification corrected several widely repeated misattributions (\ref{app:corrections}). Third, funding agencies, prize committees and early-career researchers ask, implicitly or explicitly, what kinds of careers have produced lasting benefit; an evidence-based answer is useful. Fourth, three scholars in our control list (Luenberger, Stein and Åström) died during 2026, the last on 12 November, a reminder that first-hand memory of the founding generation is disappearing \citep{stanford2026luenberger,mit2026stein,lund2026astrom,ieeecss2026astrom}.

Ranking people is nonetheless fraught. Citation counts favour large and recent fields and accumulate to those already recognised \citep{merton1968matthew}; women's contributions have been systematically under-credited \citep{rossiter1993matthew}; most modern work is produced by teams \citep{wuchty2007increasing}; and the impact of a career often appears decades after its key paper \citep{wang2013quantifying,sinatra2016quantifying}. Research-evaluation guidance therefore recommends that quantitative indicators support rather than replace qualitative judgement, that the criteria be transparent, and that composite scores be tested for robustness \citep{hicks2015leiden,nardo2008handbook,saisana2005uncertainty,saltelli2008global}. We follow that guidance throughout.

\subsection{Related work}

Table~\ref{tab:related} positions this review against earlier work. Histories of control trace the field's institutions and ideas from the governor to cybernetics and beyond \citep{mayr1970origins,bennett1979history,bennett1993history,bennett1996brief,mindell2002between,bissell2009history,kline2015cybernetics}, and \citet{astrom2014control} give an authoritative account of control as a discipline. The impact of control technology has been documented through success stories and research agendas \citep{samad2011impact,lamnabhi2017systems,murray2003future,annaswamy2023control}. Robotics is codified in handbooks and roadmaps \citep{siciliano2016springer,christensen2021roadmap,yang2018grand}, and its relation to artificial intelligence is treated by \citet{nilsson2009quest}. People-level assessments exist mainly as award citations, oral histories and citation databases; the composite indicators of \citet{ioannidis2016multiple,ioannidis2019standardized,ioannidis2020updated} are field-normalised but measure scholarly influence only. To our knowledge, no previous work ranks the scholars of control and robotics jointly, weights societal impact explicitly, and releases its scores for re-analysis.

\begin{table*}[!t]
\centering
\caption{Positioning of this review relative to earlier historical, agenda-setting and bibliometric work.}
\label{tab:related}
\footnotesize
\begin{tabular}{@{}>{\raggedright\arraybackslash}p{4.3cm} >{\raggedright\arraybackslash}p{3.3cm} >{\raggedright\arraybackslash}p{2.5cm} c c c c@{}}
\toprule
Work & Scope & Unit of analysis & Control & Robotics & Ranking & Impact weighted\\
\midrule
\citet{mayr1970origins}; \citet{bennett1979history,bennett1993history} & Control to 1955 & Ideas, institutions & \checkmark & -- & -- & --\\
\citet{mindell2002between}; \citet{kline2015cybernetics} & Feedback, cybernetics & Institutions, people & \checkmark & -- & -- & --\\
\citet{astrom2014control} & Control as a discipline & Ideas, applications & \checkmark & partly & -- & --\\
\citet{samad2011impact} & Impact of control technology & Applications & \checkmark & -- & -- & qualitative\\
\citet{lamnabhi2017systems}; \citet{annaswamy2023control} & Research agendas & Challenges & \checkmark & partly & -- & qualitative\\
\citet{siciliano2016springer} & Robotics handbook & Topics & partly & \checkmark & -- & --\\
\citet{christensen2021roadmap}; \citet{yang2018grand} & Robotics roadmaps & Challenges & -- & \checkmark & -- & qualitative\\
\citet{ioannidis2019standardized,ioannidis2020updated} & All sciences & Persons & \checkmark & \checkmark & citations & --\\
\textbf{This review} & Control and robotics, 1926-2026 & Persons & \checkmark & \checkmark & \checkmark & \checkmark\ (35\%)\\
\bottomrule
\end{tabular}
\end{table*}

\subsection{Contributions}

The contributions of this review are:
\begin{enumerate}
\item A synthesis of how ideas, subfields, schools and academic lineages evolved in control and in robotics, including corrections of common misattributions (Section~\ref{sec:evolution}, \ref{app:corrections}).
\item A map of the crossroads between the two fields: sixteen dated, illustrative pathways by which concepts crossed between the communities, and the bridge scholars who carried them (Section~\ref{sec:crossroads}).
\item A research agenda that classifies directions as most impactful, low-hanging fruit, natural progressions and wild cards, with horizons, enablers, risks and indicators (Section~\ref{sec:future}).
\item An auditable, impact-first assessment of the people behind these ideas: a documented identification and screening protocol, a source-backed scoring record for all \Npooltotal{} scored candidates, and top-100 lists for each field with life status as of 27 November 2026 (Sections~\ref{sec:method} and~\ref{sec:people}, \ref{app:lists}-\ref{app:protocol}, Supplementary Material S1).
\item A robustness analysis that perturbs both the weights and the scores, a consistency check against six award rosters, and a protocol and software for independent reassessment of the candidates near the cut-off (Section~\ref{sec:robustness}).
\end{enumerate}

\subsection{Organisation}

\begin{figure*}[!t]
\centering
\includegraphics[width=\textwidth]{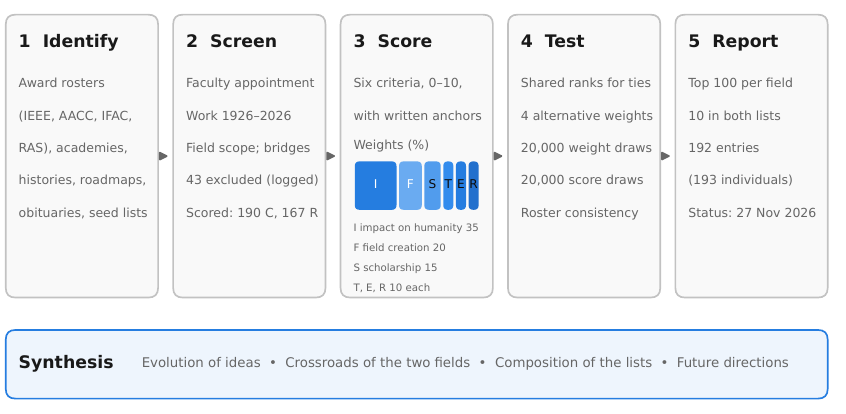}
\caption{Study design. Candidates were identified from six award rosters and a targeted search, screened by rule-based codes (every exclusion logged), scored on six anchored criteria with a source-backed evidence record, tested under perturbed weights and scores, and reported as two lists that also feed the synthesis sections. Weights are shown as a segmented bar.}
\label{fig:framework}
\end{figure*}

Section~\ref{sec:method} describes the method, summarised in Fig.~\ref{fig:framework}; the detailed protocol is in \ref{app:protocol}. Section~\ref{sec:people} introduces the scholars who lead each list; the full lists are in \ref{app:lists}. The core of the review follows: Section~\ref{sec:evolution} traces the evolution of ideas and schools, Section~\ref{sec:crossroads} the crossroads between the fields, and Section~\ref{sec:future} the research agenda. Section~\ref{sec:robustness} examines how far the lists can be trusted and what they reveal about the composition of the two communities. Section~\ref{sec:discussion} discusses lessons, ethics and limitations, and Section~\ref{sec:conclusions} concludes.

\section{Methodology}\label{sec:method}

This section states the rules; \ref{app:protocol} gives the full protocol and screening log, and Supplementary Material S1 gives the evidence behind every score.

\subsection{Scope, eligibility and field boundaries}\label{sec:scope}

\paragraph{Unit of analysis} The unit is an individual scholar. The one exception is the Denavit-Hartenberg kinematic convention, an inseparable joint contribution of a doctoral student and his adviser, which is scored as a single joint entry; counts of individuals therefore exceed counts of entries by one.

\paragraph{Eligibility} A candidate must have held, at some point in their career, a faculty appointment with a professorial or lecturer title at a university or other degree-granting institution of higher education, or a professorial post at a research institute that awards doctoral degrees (for example the institutes of the Chinese Academy of Sciences). Appointments of every type qualify: tenured, research, adjunct, affiliate, honorary, emeritus and visiting appointments, and fixed-term chairs such as the annual chairs of the Coll\`ege de France. A research post without such a title (for example at CNRS, DLR, AIST or NASA's Jet Propulsion Laboratory) does not qualify on its own. The rule is deliberately inclusive: it separates academic scholars from those whose careers were entirely industrial or governmental, not senior from junior academics. In the first version of this review the rule was stated in these terms but applied inconsistently to visiting and adjunct posts; the revised screening applies it uniformly. As a result, Kajita (professor, Chubu University), Laumond (Coll\`ege de France annual chair, 2011-2012), Cutler (distinguished adjunct professor, University of Houston), Bejczy, Baumgartner and Kitano are now scored, and Fliess, for whom no faculty title was found, is not. Every non-standard appointment is recorded with its source in S1.

\paragraph{Time window} The principal contribution must have been made between 1926 and 2026. Earlier founders, from Maxwell to Lyapunov and Minorsky, are treated as precursors (Section~\ref{sec:precursors}).

\paragraph{Field boundaries} \emph{Control systems} comprises control theory and engineering, including estimation and filtering, system identification, optimal, robust, adaptive, nonlinear, stochastic, geometric, hybrid, networked and distributed control, discrete-event systems, MPC and control-oriented optimisation, process control, guidance, navigation and flight control, control of partial differential equations, game-theoretic and learning-based control, fuzzy control, biomedical control and control education. \emph{Robotics} comprises manipulators, kinematics and dynamics, robot control, motion and task planning, robot perception and simultaneous localisation and mapping (SLAM), mobile, field, space, underwater and aerial robots, legged and humanoid robots, medical, surgical, rehabilitation and assistive robots, prostheses and exoskeletons, haptics and teleoperation, human-robot interaction, soft, micro and swarm robotics, and robot learning. Scholars whose principal contribution lies in artificial intelligence (AI), reinforcement learning as a branch of AI, mathematical optimisation, communication networks or satellite-navigation architecture are outside scope, although their influence is discussed.

\paragraph{Cross-listing rule} A scholar may be scored in both pools, but each score uses only the work relevant to that field. A scholar appears in both lists only if they qualify independently in each.

\subsection{Identification and screening}\label{sec:screening}

\paragraph{Identification} Candidates were identified from two kinds of source. The first is six award rosters, all of whose recipients were screened: the IEEE Control Systems Award (renamed the IEEE Roger W. Brockett Control Systems Award from 2026; \citealp{ieee2026csa}), the IEEE Robotics and Automation Award \citep{ieee2026raa}, the IEEE Robotics and Automation Society Pioneer Award \citep{ras2026pioneer}, the AACC Richard E. Bellman Control Heritage Award, the IFAC Giorgio Quazza Medal and the IEEE Control Systems Society Hendrik W. Bode Lecture Prize. The second is a targeted search of historical accounts and handbooks \citep{bennett1993history,astrom2014control,siciliano2016springer,annaswamy2021historical}, research agendas and roadmaps \citep{lamnabhi2017systems,annaswamy2023control,christensen2021roadmap,yang2018grand}, university and society obituaries, and citation databases used qualitatively \citep{ioannidis2019standardized}, together with the author's initial seed lists. The second kind of source is not exhaustive, so completeness is guaranteed only with respect to the rosters; \ref{app:protocol} lists the roster sources. Searches were completed and the record frozen on 27 November 2026.

\paragraph{Screening} Each identified name was either scored or excluded with one of three codes: \emph{P}, precursor (principal contribution before 1926); \emph{A}, no qualifying appointment found under the eligibility rule; \emph{B}, principal contribution outside both fields. No name was excluded on a judgement about its likely rank. Of the \NCidentified{} names identified for control, \NCpool{} were scored and \NCscreened{} excluded (\NCscreenP{} P, \NCscreenA{} A, \NCscreenB{} B); of the \NRidentified{} identified for robotics, \NRpool{} were scored and \NRscreened{} excluded (\NRscreenA{} A, \NRscreenB{} B). The screening log (Table~\ref{tab:screening}) gives each exclusion with its reason and, where an appointment was checked, the source. In the revision, \Nadded{} candidates were added to the scored pools (\NCadded{} control and \NRadded{} robotics). They comprise the names that the first version had set aside on a judgement about their likely rank, the eligibility corrections above, and the recipients of the three rosters added in the revision.

\subsection{Criteria, anchors and the evidence record}\label{sec:criteria}

Each candidate is scored on six criteria, each on an integer scale from 0 to 10 against the written anchors of Table~\ref{tab:rubric}. Impact on humanity carries the largest weight by design. It measures the breadth and depth of documented benefit to people, through deployment, safety, health, accessibility, economic value or the environment, and its durability. Credit is assessed for field-relevant work only and is shared when work was collective.

\paragraph{Era-normalised scholarly influence} Scholarly influence (S) is scored against the candidate's \emph{cohort}: the scored candidates in the same field whose signature contribution falls in the same decade. Within a cohort, signature works are compared by whether they define a named result or method taught in standard graduate courses and handbooks, by their citation standing relative to the cohort, and by whether later work is organised around them. Citation databases were consulted \citep{hirsch2005index,ioannidis2016multiple}, but raw counts were never compared across decades. The anchors are: 10, the defining work of its decade in the field; 8-9, a named result or method in standard graduate curricula; 6-7, highly cited within a subfield; 4-5, cited within a subfield; 0-3, rarely cited. As a diagnostic, the Spearman correlation between decade and S over the scored pools is \CrhoEraS{} in control and \RrhoEraS{} in robotics. The association is weak but positive, indicating that some recency effect remains.

\paragraph{Evidence record} Supplementary Material S1 lists, for every scored candidate, the six scores and the evidence for each: documented impact; field-creating and most influential works; companies, products, tools or software; textbooks, laboratories and lineages; honours, with roster awards matched automatically by name from the six identification rosters and two further rosters (the Engelberger Robotics Award and the IEEE Medal of Honor) and other honours drawn from standard biographical sources; and, where relevant, the qualifying appointment and memorial sources, each with a URL. S1 is generated from the same data files as the rankings, so the record and the lists cannot diverge.

\begin{table*}[!t]
\centering
\caption{Ranking criteria, weights and scoring anchors (integer scores 0-10). The anchors for S are applied within decade cohorts (Section~\ref{sec:criteria}).}
\label{tab:rubric}
\footnotesize
\begin{tabular}{@{}>{\raggedright\arraybackslash}p{2.6cm} c >{\raggedright\arraybackslash}p{4.5cm} >{\raggedright\arraybackslash}p{4.3cm} >{\raggedright\arraybackslash}p{3.2cm}@{}}
\toprule
Criterion & Weight & Anchor 10 & Anchor 5 & Anchor 0-2\\
\midrule
I\quad Impact on humanity & 35\% & Technology in daily use by billions, or life-critical at global scale, with a direct, documented line to the scholar's work & Documented deployment in a major sector, or theory widely used in engineering design & No documented use beyond research\\
F\quad Field or paradigm creation & 20\% & Founded a field with its own community, venues and textbooks & Founded a major method family or subfield & Incremental contributions within an existing frame\\
S\quad Scholarly influence & 15\% & The defining work of its decade in the field & Cited within a subfield (4-5); named result in graduate curricula (8-9) & Rarely cited\\
T\quad Technology transfer & 10\% & Products, companies or open tools in worldwide use & A company, product line or widely used tool & None\\
E\quad Education and community & 10\% & Canonical textbook(s) plus a large lineage of leaders, or founding of a major laboratory, institute or society & One of these at a lesser scale & None beyond normal supervision\\
R\quad Peer recognition & 10\% & Several top international prizes plus academies & Field-level society awards and fellowships & None\\
\bottomrule
\end{tabular}
\end{table*}

\subsection{Composite score and ties}\label{sec:ties}

For candidate $i$ with criterion scores $x_{ik}\in\{0,\dots,10\}$, the composite score is
\begin{equation}
S_i = 10\sum_{k=1}^{6} w_k\, x_{ik}= 3.5\,I_i + 2\,F_i + 1.5\,S_i^{\mathrm{sch}} + T_i + E_i + R_i ,
\label{eq:composite}
\end{equation}
with $\mathbf{w}=(0.35,0.20,0.15,0.10,0.10,0.10)$, so that $S_i\in[0,100]$. Candidates with equal composites are ordered lexicographically by I, F, S, T, E and R, the order of the weights, so the impact-first priority also decides ties. Candidates with identical criterion vectors cannot be separated by any rule that uses the scores; they share a rank (standard competition ranking, 1-2-2-4) and are listed alphabetically. If such a tie straddles the cut-off, all tied candidates are listed. This rule replaces the first version's tie-breaking by a robustness statistic, which, as a reviewer pointed out, inherited the input order of the data at exact ties. In the scored pools, \NCtied{} control and \NRtied{} robotics candidates belong to blocks of identical vectors; one such block, of three robotics candidates, straddles rank 100.

\subsection{Life status}

Life status is reported as of 27 November 2026, from university and society memorials, academy tributes and obituaries, cross-checked where possible \citep[e.g.][]{lund2026astrom,ieeecss2026astrom,stanford2026luenberger,mit2026stein,wonham2023memorial,ipu2018emelyanov}; S1 gives the memorial source for every death since 2015. Deceased scholars are marked \textdagger{} with their life span. A double dagger (\ddag) marks living scholars whose status is \emph{inferred} from the absence of any death report rather than confirmed by recent public activity. It is applied to everyone born in or before 1945, or whose birth year could not be verified. Life status could not be verified for some candidates outside the lists; they are flagged as unverified in the data, and the build script refuses to list any such candidate.

\subsection{Robustness analysis}\label{sec:sensitivity}

Composite indicators are only as credible as their robustness to modelling choices \citep{saisana2005uncertainty,nardo2008handbook}. We first recompute the rankings under four alternative schemes: equal weights; impact-heavy ($w_I=0.5$, others rescaled); citation-heavy ($w_S=0.35$, $w_I=0.20$, $w_F=0.15$); and no peer recognition ($w_R=0$, others renormalised). Agreement with the baseline is measured with Spearman's $\rho$ and Kendall's $\tau$ \citep{spearman1904proof,kendall1938new}. We then simulate three perturbation models, each with \NCmc{} draws: \emph{W}, weight vectors $\mathbf{w}\sim\mathrm{Dirichlet}(30\,\mathbf{w}_{\mathrm{base}})$ conditioned on $w_I=\max_k w_k$, the design constraint that impact carries the largest weight; \emph{S}, the baseline weights with every criterion score of every candidate shifted independently by $-1$, $0$ or $+1$ with probabilities 0.25, 0.5 and 0.25 (clipped to $[0,10]$), a simple model of scoring disagreement of one anchor step; and \emph{J}, both at once. In every draw, ranks are competition ranks, so candidates with identical vectors always receive identical results and the input order never matters; the build script checks both properties. For each candidate we record the rank distribution and the inclusion frequencies $W_{100}$, $S_{100}$ and $J_{100}$, the share of draws in which the candidate reaches rank 100 or better. These frequencies describe the sensitivity of the lists to the stated perturbations, with the candidate pool held fixed. They are not probabilities that a scholar belongs to a ``true'' historical top 100, and they do not capture errors of identification or of systematic bias in the scores.

\subsection{Consistency with award rosters}\label{sec:validitymethod}

We report the share of each identification roster's scored recipients that the lists include. This is a consistency check, not independent validation: the same rosters were used to identify candidates and contribute to the recognition criterion, and removing recognition from the score ($w_R=0$, reported for comparison) does not remove the selection link.

\subsection{Independent reassessment}\label{sec:reassess}

Membership near the cut-off depends on the scores most of all. The candidates at baseline ranks 81-120 in each field (\Nreassess{} candidates) are therefore subject to independent reassessment. At least two assessors per field, who are not authors, hold a doctorate in control or robotics, and declare no advisor-student or co-authorship relation with the candidates they score, score the six criteria independently from blind sheets that carry the S1 evidence but not the authors' scores or ranks. Agreement is measured with quadratically weighted Cohen's $\kappa$ per criterion and the intraclass correlation of the composite, and the lists are recomputed with the averaged scores under the same tie rule. The sheets and the analysis script are released (S1, Section~6).

\subsection{Use of AI tools in the research process}\label{sec:aiuse}

The author used a generative AI system, Claude (Anthropic; Claude Opus model 5.5, accessed through the Claude application in November 2026), at the following stages of the research. (i) Identification and verification: running web searches and retrieving sources for candidate identification, roster transcription, appointment checks and life status. Every eligibility decision and death date in S1 cites the source that was retrieved. (ii) Scoring: proposing initial criterion scores and short evidence statements against the anchors of Table~\ref{tab:rubric}. The author reviewed each score and evidence statement and is responsible for them; the independent reassessment of Section~\ref{sec:reassess} is designed in part to test them. (iii) Software: writing the Python code that computes the rankings, robustness analyses, tables and the scoring record. (iv) Figures: all figures are data visualisations or schematic diagrams rendered deterministically by that code (matplotlib) from the released data; no generative image model was used. The code and data are released so that every number, table and figure can be regenerated and checked.

\subsection{Threats to validity}\label{sec:threats}

Six threats deserve explicit mention. (i) \emph{Judgement}: the scores are structured judgements, not measurements; the anchors, evidence record, score perturbations and independent reassessment mitigate but do not remove this. (ii) \emph{Cumulative advantage}: fame begets citations and awards \citep{merton1968matthew}; within-cohort scoring of S and the low weight on recognition limit the effect, but the weak positive correlation between decade and S shows that it is not eliminated. (iii) \emph{Under-credited groups}: women's contributions have historically been under-recognised \citep{rossiter1993matthew}, and scholarship published in Russian, Japanese or Chinese is less visible in English-language sources; we deliberately searched those schools, but residual bias is likely. (iv) \emph{Attribution}: deployed technology is the work of teams \citep{wuchty2007increasing}; we credit academics only for documented contributions and list industrial originators separately. (v) \emph{Recency}: the impact of recent work is still unfolding \citep{wang2013quantifying}, which disadvantages early-career scholars; we therefore make no ranking of rising stars. (vi) \emph{Boundaries and identification}: the eligibility and field-scope rules exclude some influential people (\ref{app:context}), and identification beyond the six rosters depends on a targeted, non-exhaustive search.

\section{The scholars who shaped the two fields}\label{sec:people}

\begin{figure*}[!t]
\centering
\includegraphics[width=\textwidth]{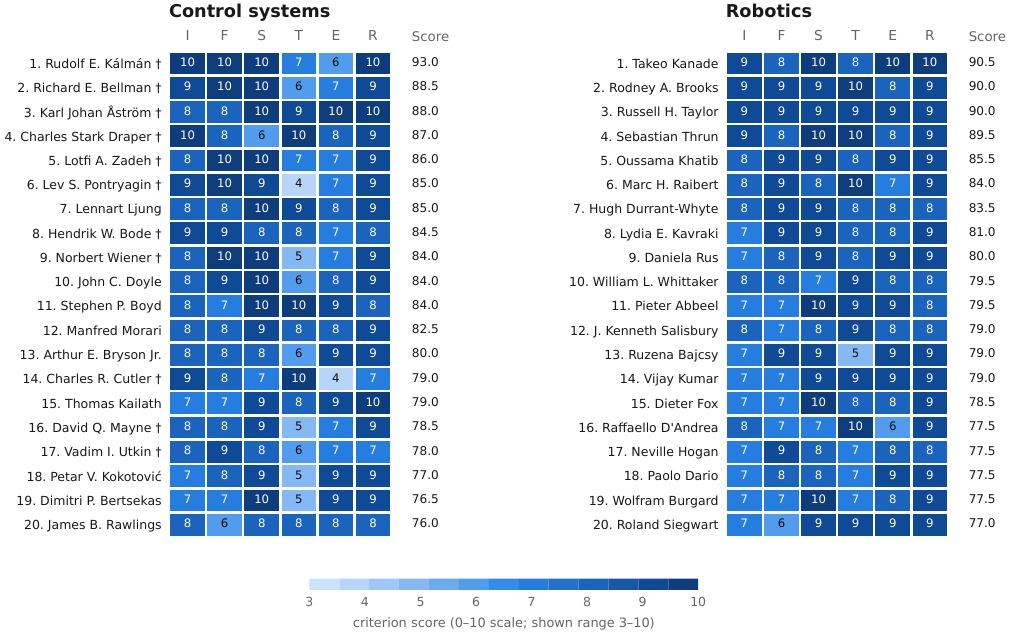}
\caption{Criterion profiles of the twenty highest-ranked scholars in each field (I impact on humanity, F field creation, S scholarly influence, T technology transfer, E education and community, R peer recognition). Cells show integer scores; the right-hand column is the composite score of Eq.~\eqref{eq:composite}. \textdagger{} deceased.}
\label{fig:profiles}
\end{figure*}

This section introduces the scholars at the head of each list; the complete lists, with institutions, signature contributions, impact statements and scores, are Tables~\ref{tab:control100} and~\ref{tab:robotics100} in \ref{app:lists}, and the evidence for each score is in S1. Ranks are given in parentheses. Fig.~\ref{fig:profiles} shows the criterion profiles of the leading twenty in each field.

\subsection{Control systems}\label{sec:control}

The control list has composite scores from \Cmax{} to \Ccutoff{}, with a median of \Cmedian{}. It contains \NCdead{} deceased and \NCliving{} living scholars, \NCflag{} of the latter with inferred status.

\paragraph{Architects of the state-space revolution} Within roughly five years, three ideas turned control from a toolkit into a mathematical discipline. \textbf{Rudolf E.~Kálmán} (1) formulated recursive optimal filtering in state space \citep{kalman1960new,kalman1961new} and, with it, controllability, observability and minimal realisation. The filter was adapted at NASA Ames for Apollo circumlunar navigation \citep{mcgee1985discovery} and has since become the estimation engine of aerospace, navigation receivers and tracking systems \citep{grewal2010applications}. \textbf{Richard E.~Bellman} (2) introduced dynamic programming and the principle of optimality \citep{bellman1952theory,bellman1957dynamic}, the common root of optimal control, operations research and reinforcement learning \citep{bertsekas1996neuro,sutton2018reinforcement}. \textbf{Lev S.~Pontryagin} (6), blind from adolescence, led the Moscow group that proved the maximum principle \citep{pontryagin1962mathematical}, the necessary conditions behind trajectory optimisation in aerospace. \textbf{Arthur E.~Bryson~Jr.} (13) made optimal control usable: \emph{Applied Optimal Control} \citep{bryson1969applied}, written with Yu-Chi Ho, taught gradient-based trajectory optimisation and adjoint methods to a generation of aerospace engineers. \textbf{Thomas Kailath} (15) built the linear-systems and fast-estimation theory that links control with communications and signal processing \citep{kailath1980linear}.

\paragraph{Feedback before state space} Three earlier figures established feedback as engineering practice. \textbf{Charles Stark Draper} (4) created inertial guidance and navigation at the MIT Instrumentation Laboratory, from wartime gunsights to the Apollo guidance, navigation and control system \citep{draper1960inertial,hall1996journey,mindell2008digital}; his impact score is the highest possible, his scholarly-influence score modest. \textbf{Hendrik W.~Bode} (8) gave feedback design its frequency-domain language, including the gain-phase relations and the sensitivity integral \citep{bode1945network}, and spent his final years as a Harvard professor. \textbf{Norbert Wiener} (9) derived optimal linear filtering and framed feedback as a principle spanning machines, organisms and society \citep{wiener1948cybernetics,wiener1949extrapolation,kline2015cybernetics}.

\paragraph{Adaptation, identification and approximate reasoning} \textbf{Karl Johan Åström} (3), who died on 12 November 2026, founded the Department of Automatic Control at Lund University, introduced the belief-state formulation of partially observed control \citep{astrom1965optimal}, the self-tuning regulator \citep{astrom1973selftuning}, relay auto-tuning of PID controllers \citep{astrom1984automatic} and event-based (Lebesgue) sampling \citep{astrom2002comparison}, and wrote textbooks that still define the field \citep{astrom2008feedback}. \textbf{Lennart Ljung} (7), his student, built the prediction-error framework of system identification and its software \citep{ljung1999system}, a standard route by which industrial models are obtained from data. \textbf{Lotfi A.~Zadeh} (5) introduced fuzzy sets \citep{zadeh1965fuzzy}. Fuzzy controllers, first realised by \citet{mamdani1974application}, spread through consumer products and industry. Earlier, Zadeh co-wrote the textbook that brought the state-space approach to engineers \citep{zadeh1963linear}.

\paragraph{Robustness and optimisation} \textbf{John C.~Doyle} (10) showed that linear-quadratic-Gaussian designs carry no guaranteed robustness margins \citep{doyle1978guaranteed}, introduced the structured singular value \citep{doyle1982analysis} and, with Glover, Khargonekar and Francis, gave the state-space solution of $\mathcal{H}_\infty$ control \citep{doyle1989state}. \textbf{Stephen P.~Boyd} (11) recast control design as convex optimisation \citep{boyd1994linear,boyd2004convex} and produced software (CVX, CVXPY, CVXGEN) that moved optimisation into embedded systems \citep{grant2014cvx,diamond2016cvxpy,mattingley2012cvxgen}; SpaceX reports using CVXGEN-generated code for rocket landing \citep{blackmore2016autonomous}. \textbf{Manfred Morari} (12) developed internal model control \citep{garcia1982internal} and the robust, hybrid and explicit forms of MPC \citep{morari1989robust,bemporad1999control,bemporad2002explicit}. \textbf{Charles R.~Cutler} (14), who died in 2020, developed dynamic matrix control at Shell \citep{cutler1980dynamic}, the first widely deployed industrial MPC, founded DMC Corporation (later part of AspenTech) and was a distinguished adjunct professor at the University of Houston; industrial predictive control now runs throughout refining and petrochemicals \citep{qin2003survey}. He enters the list because the revised screening applies the eligibility rule to adjunct appointments consistently. \textbf{David Q.~Mayne} (16) co-invented differential dynamic programming \citep{mayne1966second,jacobson1970differential}, an early Monte Carlo filter \citep{handschin1969monte} and the stability theory of constrained MPC \citep{mayne2000constrained}. \textbf{James B.~Rawlings} (20) consolidated that theory, together with moving-horizon estimation, into the reference text used in industry and academia \citep{rawlings2017model}. \textbf{Dimitri P.~Bertsekas} (19) unified dynamic programming, optimisation and approximate (neuro-dynamic) programming \citep{bertsekas1996neuro,bertsekas2019reinforcement}.

\paragraph{Nonlinear and variable-structure control} \textbf{Vadim I.~Utkin} (17) developed, with Stanislav Emelyanov (38), the theory of sliding modes and equivalent control \citep{emelyanov1967variable,utkin1977variable,utkin1992sliding}, which is now widely used in electric drives and power converters. The two shared the 1972 Lenin Prize for this work \citep{ipu2018emelyanov}. \textbf{Petar V.~Kokotović} (18) systematised singular-perturbation methods \citep{kokotovic1986singular} and, with Kanellakopoulos and Krstić, backstepping \citep{krstic1995nonlinear}. He also trained an unusually productive lineage (Section~\ref{sec:schools}).

\paragraph{Beyond the leading twenty} Ranks 21-60 are dominated by the architects of modern theory: optimal control, differential games and discrete-event systems (Ho, 21), adaptive control (Goodwin, Narendra, Ioannou, and Clarke's generalised predictive control, \citealp{clarke1987generalized}), optimal filtering and observers (Anderson, Luenberger), stochastic systems and dynamic games (Varaiya, Başar), nonlinear and geometric control (Isidori, Sontag, Brockett, Wonham, Krstić, Sastry, Khalil), robust and multivariable control (Glover, Zames, Rosenbrock, Athans, Stein), passivity and dissipativity (Willems), tracking (Bar-Shalom) and the Columbia school of sampled-data control led by Ragazzini. Applied impact lifts several engineering-oriented scholars into this band. Gordon S.~Brown's MIT Servomechanisms Laboratory originated numerically controlled machining \citep{brown1948principles}. Malcolm C.~Smith's inerter moved from network synthesis into racing-car and railway suspensions \citep{smith2002synthesis}. Tomizuka's zero-phase-error tracking and repetitive control underlie precision servo systems \citep{tomizuka1987zero}. Gilbert's distance algorithm became a standard collision check in robotics and physics simulation \citep{gilbert1988fast}. Closed-loop insulin delivery is represented by Kovatchev (43), whose simulator and algorithms underpin commercial systems \citep{kovatchev2009insilico}, by Hovorka (53) \citep{hovorka2004nonlinear} and by Doyle~III \citep{grosman2010zone}. The band also includes Isermann for model-based fault diagnosis \citep{isermann2005model}, Sugeno for fuzzy modelling and control \citep{takagi1985fuzzy}, and Bemporad and Skogestad for embedded predictive control and process-control practice \citep{bemporad2002explicit,skogestad2003simple}. Ranks 61-100 include van der Schaft, Francis and Kushner (sharing rank 61), the Soviet school (Andronov, Tsypkin, Yakubovich), control of partial differential equations (Lions), biomedical modelling (Cobelli), industrial process automation in China (Chai, \citealp{chai2014optimal}), process systems engineering (Biegler, Ray), the robust servomechanism problem (Davison), parameter-space robust control (Ackermann), irrigation-network automation (Mareels), networked and distributed systems (P.~R.~Kumar, Tsitsiklis, Johansson), and the safety, optimisation and learning generation (Tomlin, Pappas, Ames, Tedrake, Borrelli, Dörfler). The twenty candidates immediately below the cut-off are listed in Table~\ref{tab:hm}.

\subsection{Robotics}\label{sec:robotics}

The robotics list has composite scores from \Rmax{} to \Rcutoff{}, with a median of \Rmedian{}. Because three candidates with identical criterion vectors share rank 100, it has \NRtop{} entries. Robotics is the younger academic field, and the list contains \NRdead{} deceased and \NRliving{} living scholars, \NRflag{} of the latter with inferred status.

\paragraph{Seeing and acting} \textbf{Takeo Kanade} (1) ranks first because his work spans perception, mechanism and autonomy, each with documented reach. The Lucas-Kanade registration method \citep{lucas1981iterative} is a building block of motion tracking; the direct-drive arm \citep{asada1983design} removed gearing from manipulator design; and the Navlab vehicles \citep{thorpe1988vision} were among the first self-driving cars. As director of the Carnegie Mellon Robotics Institute he also built one of the field's largest training grounds. \textbf{Ruzena Bajcsy} (13) reframed perception as an active, purposive process \citep{bajcsy1988active}, founded the GRASP Laboratory, and introduced elastic matching, which underlies anatomical atlas registration in medical imaging \citep{bajcsy1989multiresolution}.

\paragraph{Architectures and autonomy} \textbf{Rodney A.~Brooks} (2) replaced the sense-plan-act pipeline with the subsumption architecture \citep{brooks1986robust,brooks1991intelligence}. As a co-founder of iRobot he helped bring robots into tens of millions of homes \citep{jones2006robots}. \textbf{Sebastian Thrun} (4), with Wolfram Burgard (19) and Dieter Fox (15), turned uncertainty into the organising principle of mobile robotics \citep{thrun2005probabilistic}. Their Monte Carlo localisation \citep{dellaert1999monte}, museum tour-guide robots \citep{burgard1999experiences}, grid-mapping methods \citep{grisetti2007improved} and dynamic-window navigation \citep{fox1997dynamic} became standard. Thrun's Stanley won the 2005 DARPA Grand Challenge \citep{thrun2006stanley}, and he went on to lead Google's self-driving car project, which seeded the autonomous-driving industry. \textbf{Hugh Durrant-Whyte} (7) established SLAM and decentralised data fusion \citep{leonard1991mobile,durrantwhyte2006simultaneous} and translated them into automated container terminals and mining fleets \citep{durrantwhyte2007autonomous}. \textbf{William L.~Whittaker} (10) founded field robotics, from robots for the Three Mile Island clean-up to Boss, the 2007 Urban Challenge winner \citep{urmson2008autonomous}. \textbf{Roland Siegwart} (20) combined mobile- and aerial-robot research \citep{siegwart2011introduction} with an exceptional record of spin-off companies.

\paragraph{Manipulation, interaction and haptics} \textbf{Oussama Khatib} (5) introduced artificial potential fields \citep{khatib1985real,khatib1986real} and the operational-space formulation \citep{khatib1987unified}, the basis of whole-body and human-safe manipulation. His OceanOne humanoid extended human presence to the deep sea \citep{khatib2016ocean}. \textbf{J.~Kenneth Salisbury} (12) designed articulated hands \citep{salisbury1982articulated} and the PHANToM haptic interface \citep{massie1994phantom}, a device that put force feedback into laboratories and clinics. \textbf{Neville Hogan} (17) formulated impedance control \citep{hogan1985impedance} and co-developed the MIT-Manus rehabilitation robot \citep{krebs1998robot}. Its successor was evaluated in a multicentre randomised trial in stroke survivors \citep{lo2010robot}.

\paragraph{Planning} \textbf{Lydia E.~Kavraki} (8) introduced probabilistic roadmaps \citep{kavraki1996probabilistic}, which made high-dimensional motion planning practical, and led the Open Motion Planning Library \citep{sucan2012open}; her methods also serve molecular biology. The configuration-space foundations on which this work rests are due to Lozano-Pérez (24), and the field's first comprehensive text to Latombe (29) \citep{lozanoperez1983spatial,latombe1991robot}.

\paragraph{Legs, wings and fleets} \textbf{Marc H.~Raibert} (6) showed that dynamic balance, not static stability, is the key to legged locomotion \citep{raibert1986legged}. He founded Boston Dynamics, whose legged robots are now used commercially \citep{raibert2008bigdog}. \textbf{Vijay Kumar} (14) created agile aerial robots and swarms through minimum-snap trajectory design \citep{mellinger2011minimum,kushleyev2013towards}. \textbf{Raffaello D'Andrea} (16), a bridge scholar, co-founded Kiva Systems, whose coordinated mobile robots \citep{wurman2008coordinating} became the model for e-commerce fulfilment after their acquisition by Amazon in 2012.

\paragraph{Medicine, biorobotics and learning} \textbf{Russell H.~Taylor} (3) is the principal architect of computer-integrated surgery, from ROBODOC \citep{taylor1994image} to cooperative ``steady-hand'' systems and an engineering research centre that trained much of the field \citep{taylor2003medical}. \textbf{Paolo Dario} (18) built biorobotics as a discipline and an institute \citep{dario1996robotics}. \textbf{Daniela Rus} (9) advanced self-reconfiguring, distributed and soft robots \citep{rus2001crystalline,rus2015design} and leads MIT's largest computing laboratory. \textbf{Pieter Abbeel} (11) brought apprenticeship and deep reinforcement learning to robots \citep{abbeel2004apprenticeship,abbeel2010autonomous,levine2016end} and to industry.

\paragraph{Beyond the leading twenty} Ranks 21-60 combine four groups. The first are the founders of the mechanics, kinematics and control of manipulation: Lozano-Pérez, Vukobratović (the zero-moment point), Latombe, Paul, Mason, Roth, Bicchi and Slotine. The second are the creators of humanoid, legged and bio-inspired machines: Hirose, Kato, Hutter, Kazerooni, Wood and Floreano. The third are scholars whose devices reached patients, factories, roads and consumers. They include Herr and Sankai (powered prostheses and exoskeletons), Makino (the SCARA arm), Clavel (the Delta robot), Jacobsen (the Utah Arm) and Hirzinger (lightweight torque-controlled arms). They also include Haddadin (collision safety), Okamura and Yang (haptics and surgical robotics), Riener (gait rehabilitation), Nelson (medical microrobots) and Dickmanns (vision-based autonomous road vehicles, \citealp{dickmanns1987autonomous}). Zexiang Li combined a standard robotics text with the academic entrepreneurship behind DJI's drones. LaValle introduced rapidly exploring random trees \citep{lavalle2001randomized}, and Goldberg brought cloud robotics and learned grasping to logistics \citep{mahler2017dexnet}. The fourth group spans learning (Levine, Finn), SLAM (Davison, Dellaert, Leonard), social and assistive robotics (Breazeal, Matarić) and supervisory control of telerobots (Sheridan, \citealp{sheridan1992telerobotics}). Ranks 61-100 add industrial-mechanism and assembly science (Whitney, Gosselin, Freudenstein, the Denavit-Hartenberg convention) and Japanese humanoid and telexistence research (Kajita, whose preview control of the zero-moment point is used across humanoid robots, \citealp{kajita2003biped}; Nakamura, Tachi, Ishiguro and Mori, following Inoue at rank 60). They also add nonholonomic motion planning and its industrial software (Laumond), soft and bio-inspired robotics (Laschi, with Ijspeert and Full among the near-misses), surgical robotics (Davies) and robot learning and navigation (Schaal, Billard, Scaramuzza). The list closes with Ames, Ishiguro and Sitti, whose identical criterion vectors give them a shared rank of 100. Table~\ref{tab:hm} lists the candidates immediately below them, several of whom enter under alternative weightings (Section~\ref{sec:robustness}).

\section{Evolution of ideas and subfields}\label{sec:evolution}

\begin{figure*}[p]
\centering
\includegraphics[width=0.97\textwidth]{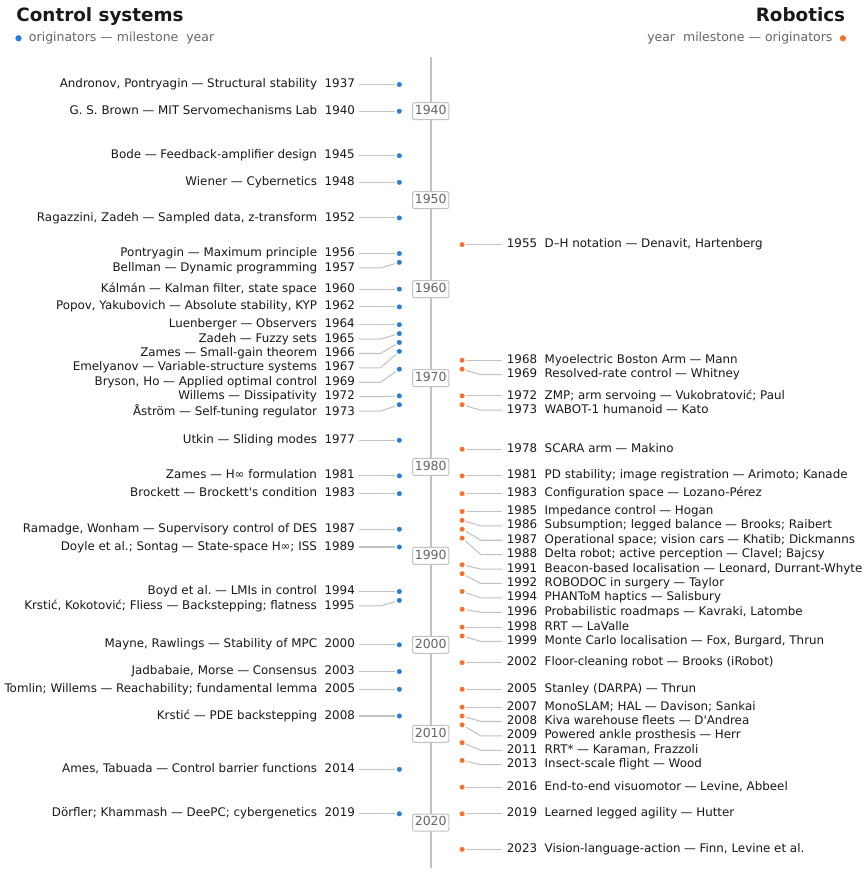}
\caption{Milestones in control systems (left) and robotics (right), 1937-2023, with their originators. Leader lines connect each milestone to its year on the central axis. Industrial milestones without an academic originator, such as the Unimate (1961), are discussed in the text and in Appendix~B.}
\label{fig:timeline}
\end{figure*}

Fig.~\ref{fig:timeline} places the milestones of both fields on a common time axis. This section narrates that evolution and the schools behind it.

\subsection{Precursors before 1926}\label{sec:precursors}

The analysis of feedback began with Maxwell's study of governors \citep{maxwell1868governors}, followed by Vyshnegradsky's design rules \citep{vyshnegradsky1876}, the stability criteria of Routh and Hurwitz \citep{routh1877treatise,hurwitz1895bedingungen}, Lyapunov's theory of the stability of motion \citep{lyapunov1992general} and Minorsky's three-term ship-steering controller \citep{minorsky1922directional}. Robotics inherited its name from \v{C}apek's play \emph{R.U.R.} \citep{capek1920rur}, its kinematics from nineteenth-century mechanism theory, and its first industrial embodiment from Devol's patent and Engelberger's Unimate \citep{devol1961programmed,engelberger1980robotics}. The industrial origin of the feedback amplifier \citep{black1934stabilized} and of the Nyquist criterion \citep{nyquist1932regeneration} explains why several founding figures of control appear in \ref{app:context} rather than in the ranked list.

\subsection{Control systems}

\paragraph{Servomechanisms and the frequency domain (1930s-1950s)} Wartime fire control fused electronics, mechanics and mathematics. Bode's design theory, the MIT Servomechanisms and Radiation Laboratories, Wiener's filtering, the root locus \citep{evans1948graphical} and tuning rules \citep{ziegler1942optimum} produced the classical toolkit. In parallel, Andronov's Gorky school and the Moscow Institute of Automation and Remote Control developed nonlinear oscillation and stability theory \citep{andronov1937systemes,andronov1966theory}. Sampled-data control followed at Columbia, where Ragazzini's students included Kálmán, Zadeh, Jury and Franklin \citep{ragazzini1952analysis,jury1964theory}.

\paragraph{The state-space revolution (1956-1965)} The maximum principle, dynamic programming and the Kalman filter appeared within a decade. The International Federation of Automatic Control was founded in 1957 and held its first World Congress in Moscow in 1960. Observers \citep{luenberger1964observing}, pole assignment \citep{wonham1967pole} and the absolute-stability results of Popov and Yakubovich \citep{popov1962absolute,yakubovich1962solution} completed a coherent linear theory, codified in textbooks \citep{zadeh1963linear,athans1966optimal,bryson1969applied,brockett1970finite}.

\paragraph{Stochastic, adaptive and data-driven modelling (1960s-1990s)} Stochastic control and filtering matured \citep{kushner1967stochastic,fleming1975deterministic,zakai1969optimal,anderson1979optimal}, Lund and Linköping made adaptation and identification industrial \citep{astrom1973selftuning,ljung1999system}, and stability proofs for adaptive control were established \citep{morse1980global,goodwin1980discrete,narendra1989stable,ioannou1996robust}.

\paragraph{Multivariable and robust control (1970s-1990s)} Multivariable frequency-domain methods \citep{rosenbrock1970state,macfarlane1977generalized} were followed by the realisation that optimal designs could be fragile \citep{doyle1978guaranteed}, by the small-gain and $\mathcal{H}_\infty$ frameworks \citep{zames1966input,zames1981feedback}, by the parametrisation of all stabilising controllers \citep{youla1976modern,kucera1979discrete,vidyasagar1985control} and by the state-space solution \citep{doyle1989state}. Robust control entered flight control through LQG/LTR and loop shaping \citep{doyle1981multivariable,mcfarlane1992loop}.

\paragraph{Nonlinear and geometric control (1970s-2000s)} Differential geometry brought controllability, feedback linearisation and zero dynamics \citep{sussmann1972controllability,hermann1977nonlinear,isidori1981nonlinear,hunt1983global,isidori1995nonlinear}. Constructive designs followed in the form of input-to-state stability, control Lyapunov functions, backstepping, sliding modes and flatness \citep{sontag1989smooth,artstein1983stabilization,krstic1995nonlinear,utkin1992sliding,fliess1995flatness}. Brockett's condition \citep{brockett1983asymptotic} showed what smooth feedback cannot do, with direct consequences for robots (Section~\ref{sec:crossroads}).

\paragraph{Discrete-event, hybrid, networked and distributed systems (1980s-2010s)} Supervisory control \citep{ramadge1987supervisory}, hybrid systems \citep{tomlin1998conflict,alur2000discrete,goebel2012hybrid}, networked and event-triggered control \citep{hespanha2007survey,astrom2002comparison,tabuada2007event,heemels2012introduction} and consensus \citep{jadbabaie2003coordination,olfati2004consensus} extended control to software, networks and fleets. A recent review in this journal shows the event-triggered line now reaching autonomous robots \citep{yu2026bibliographical}.

\paragraph{Optimisation-based control (1978-)} MPC began in industry \citep{richalet1978model,cutler1980dynamic} and acquired its theory in academia \citep{garcia1989model,mayne2000constrained,rawlings2017model}. Convex optimisation \citep{boyd1994linear} and explicit and embedded MPC \citep{bemporad2002explicit,mattingley2012cvxgen} then moved it into fast, safety-critical systems \citep{mayne2014model}.

\paragraph{Data-driven and learning-based control (2005-)} The behavioural ``fundamental lemma'' \citep{willems2005note} grounded data-enabled predictive control \citep{coulson2019data,berberich2021data,markovsky2021behavioral}. Reinforcement learning was reinterpreted as approximate dynamic programming \citep{bertsekas1996neuro,recht2019tour}, and learning-based MPC and safety filters address its certification \citep{hewing2020learning,brunke2022safe}.

\paragraph{Control in biology and medicine (2000-)} Feedback now regulates insulin delivery for people with type~1 diabetes \citep{hovorka2004nonlinear,grosman2010zone,brown2019six}, validated in silico on an FDA-accepted simulator \citep{kovatchev2009insilico}, and is engineered inside living cells \citep{briat2016antithetic,aoki2019universal}.

\subsection{Robotics}

\paragraph{Kinematics and the first arms (1955-1975)} The Denavit-Hartenberg notation \citep{denavit1955kinematic} and Freudenstein's kinematic synthesis \citep{freudenstein1955approximate} supplied the mathematics; Stanford and MIT supplied the first computer-controlled arms, inverse kinematics and programming \citep{pieper1969kinematics,paul1972modelling,whitney1969resolved}; and Waseda built WABOT-1, the first full-scale anthropomorphic robot \citep{kato1974information}. Vukobratović's zero-moment point \citep{vukobratovic1972stability} and powered exoskeletons \citep{vukobratovic1974development} laid foundations for biped and wearable robotics.

\paragraph{Dynamics, control and interaction (1975-1990)} Manipulator control drew directly on control theory, through computed torque, PD stability \citep{takegaki1981new}, compliance and impedance \citep{mason1981compliance,hogan1985impedance}, operational space \citep{khatib1987unified} and adaptive control \citep{slotine1987adaptive,craig1987adaptive}, while industrial practice adopted the SCARA and Delta architectures \citep{makino2014development,clavel1988delta}.

\paragraph{Motion planning (1979-)} Configuration space \citep{lozanoperez1979algorithm,lozanoperez1983spatial} and complexity results \citep{canny1988complexity} gave way to sampling-based planners \citep{kavraki1996probabilistic,lavalle2001randomized,karaman2011sampling}, whose variants are still being refined \citep{venu2026optimized}, and to integrated task and motion planning \citep{kaelbling2011hierarchical,garrett2021integrated}.

\paragraph{Behaviour-based and probabilistic robotics (1986-2010)} The subsumption architecture \citep{brooks1986robust} and behaviour-based systems \citep{arkin1998behavior,mataric1992integration} challenged deliberative AI. Probabilistic state estimation, from occupancy grids \citep{moravec1985high} and stochastic mapping \citep{smith1986representation} through localisation \citep{leonard1991mobile,dellaert1999monte} and SLAM \citep{durrantwhyte2006simultaneous,davison2007monoslam,murartal2015orb,cadena2016past}, became the backbone of mobile autonomy and of the DARPA challenges \citep{thrun2006stanley,buehler2009darpa,krotkov2017darpa}.

\paragraph{Legged and humanoid robots} From WABOT-1 and the zero-moment point to Raibert's hopping machines \citep{raibert1986legged}, hybrid zero dynamics \citep{westervelt2003hybrid}, preview control of the zero-moment point \citep{kajita2003biped} and learned locomotion \citep{hwangbo2019learning,lee2020learning,miki2022learning}, legged robotics moved from laboratory curiosity to commercial inspection platforms.

\paragraph{Medical, rehabilitation and assistive robotics} The Boston Arm \citep{mit2006mann}, the Utah Arm \citep{jacobsen1982development}, ROBODOC \citep{taylor1994image}, active surgical robots \citep{davies1991development}, rehabilitation robots \citep{krebs1998robot,riener2005patient}, exoskeletons \citep{kazerooni2005control,suzuki2007intention} and powered prostheses \citep{au2009powered} trace a sixty-year path from feasibility to clinical evidence \citep{lo2010robot}, now complemented by bio-inspired controllers for wearable robots \citep{almeida2025bio}.

\paragraph{Field, aerial, soft, micro and swarm robotics} Field robotics \citep{urmson2008autonomous}, aerial robotics \citep{mellinger2011minimum,floreano2015science,kaufmann2023champion}, soft robotics \citep{laschi2012soft,rus2015design,hawkes2017soft}, micro-robotics \citep{nelson2010microrobots,sitti2015biomedical} and swarms \citep{rubenstein2014programmable,brambilla2013swarm} broadened what counts as a robot.

\paragraph{Robot learning and foundation models (2015-)} End-to-end visuomotor learning \citep{levine2016end,levine2018learning}, meta-learning \citep{finn2017model}, simulation with domain randomisation \citep{todorov2012mujoco,tobin2017domain} and, since 2023, vision-language-action models trained on pooled multi-robot data \citep{brohan2023rt2,openx2024,kim2024openvla,black2024pi0} have made generalist manipulation a central research goal.

\subsection{Schools and lineages}\label{sec:schools}

\begin{figure*}[!t]
\centering
\includegraphics[width=\textwidth]{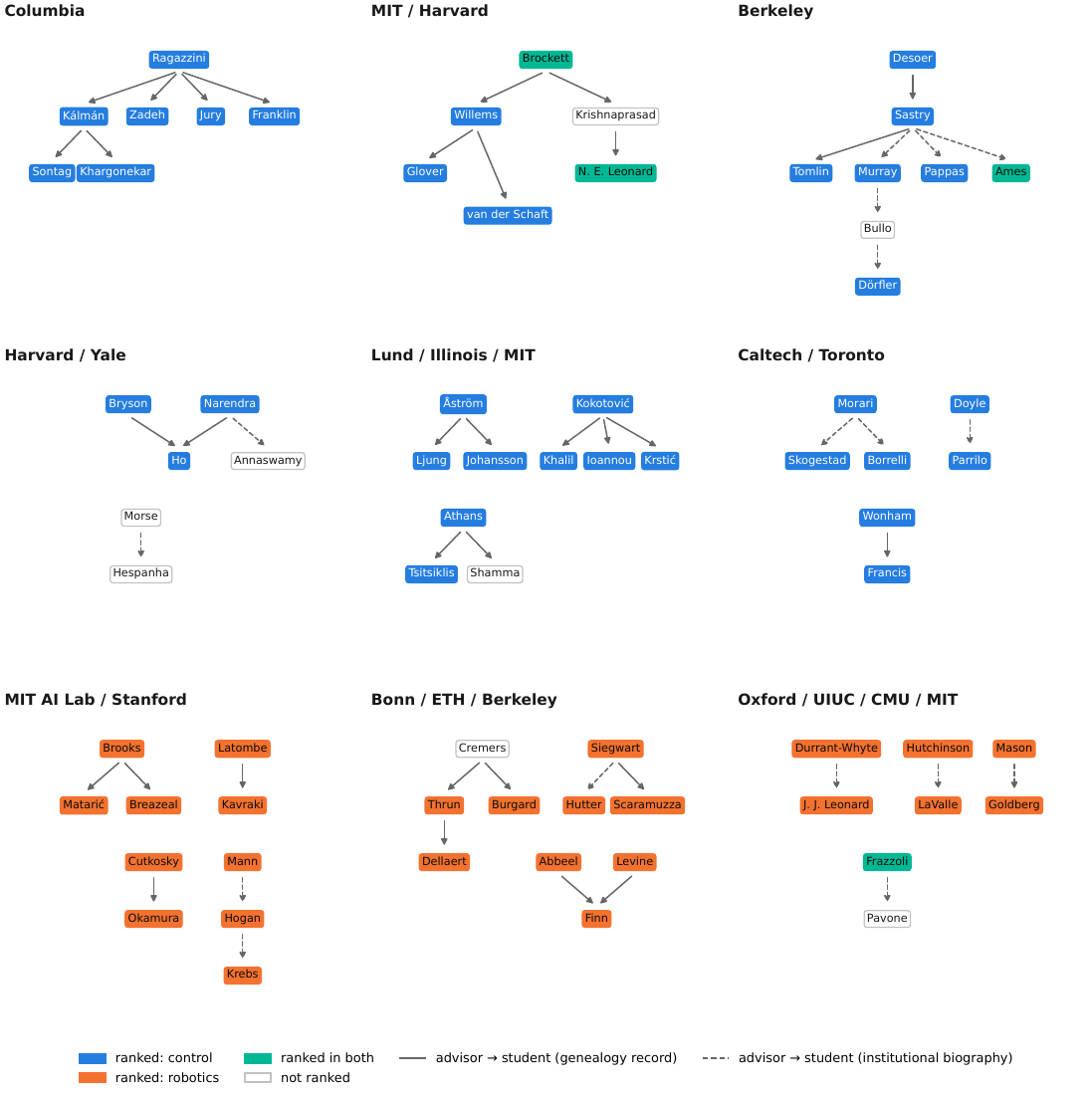}
\caption{Selected academic lineages among the ranked scholars. Solid arrows: advisor-student links documented in public genealogy records; dashed arrows: links documented in institutional biographies. Fill colour indicates the list(s) in which a scholar is ranked. The figure is illustrative, not exhaustive.}
\label{fig:genealogy}
\end{figure*}

Fig.~\ref{fig:genealogy} shows selected advisor-student links among the ranked scholars, compiled from genealogy records and institutional biographies. Four patterns stand out. First, schools compound. Ragazzini's Columbia group produced Kálmán, Zadeh, Jury and Franklin, and Kálmán in turn advised Sontag and Khargonekar. Brockett advised Willems, whose students include Glover and van der Schaft. Desoer's student Sastry trained Tomlin, Murray, Pappas and Ames, and Murray's student Bullo trained Dörfler. Second, the founding generation of probabilistic robotics shared a single origin: Thrun and Burgard both completed doctorates in Cremers' group at Bonn. Third, lineages cross the field boundary: Brockett's line reaches Naomi Leonard's underwater gliders, and Sastry's line reaches Ames's bipeds and exoskeletons. Fourth, the hubs are few: MIT, Stanford, Berkeley, Caltech, Carnegie Mellon, ETH Zürich, Lund, Illinois and Tokyo account for a large share of both lists.

\section{Crossroads between robotics and control}\label{sec:crossroads}

\subsection{Concepts that migrated}\label{sec:migration}

\begin{figure*}[!t]
\centering
\includegraphics[width=0.92\textwidth]{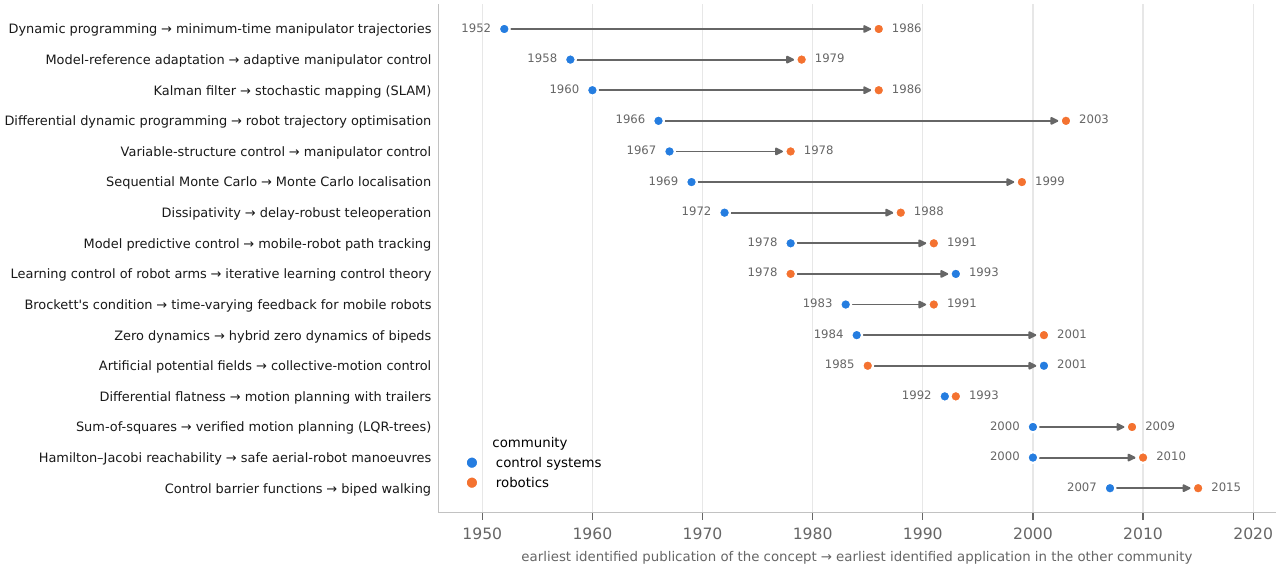}
\caption{Illustrative concept-migration pathways. Each row links the earliest identified publication of a concept (tail) to the earliest identified application in the other community (head); colour indicates the community. Dates follow the rule of Section~\ref{sec:migration}; references are in Table~\ref{tab:migration}.}
\label{fig:migration}
\end{figure*}

Table~\ref{tab:migration} and Fig.~\ref{fig:migration} document sixteen pathways by which a concept that originated in one community was taken up by the other. The pathways are illustrative, not a sample: they were chosen because each is well documented and each connects scholars in the lists. All rows are dated by one rule. The \emph{origin} is the earliest archival publication of the concept that we identified, and the \emph{adoption} is the earliest archival publication we identified that applies the concept to the systems or problems of the other community. Both dates come from a targeted search, so earlier instances may exist, and individual lags are indicative only. The first version of this table dated several rows inconsistently, for example by landmark later papers rather than first applications; a reviewer identified three such cases, and all sixteen rows were re-dated by the rule above.

Three observations survive this caveat. (i) Fourteen of the sixteen pathways run from control to robotics. Estimation, dynamic programming, adaptation, variable-structure control, passivity, predictive control, flatness, sum-of-squares verification, reachability and barrier functions all crossed in that direction, usually because manipulators, mobile robots and legged machines offered demanding nonlinear test beds. Because the pathways were selected rather than sampled, this is a description of well-documented cases, not an estimate of the overall balance of exchange. (ii) Influence also flowed back. Learning control was formulated for robot arms \citep{uchiyama1978formation,arimoto1984bettering} before it became a general theory of iterative learning control \citep{moore1993iterative,bristow2006survey}; artificial potential fields \citep{khatib1985real} reappeared as the virtual potentials of collective-motion control \citep{leonard2001virtual}; and nonsmooth contact and impact problems from robotics continue to pose open control questions \citep{brogliato2023modeling}. (iii) Some concepts were born at the crossroads rather than carried across it. Hybrid zero dynamics, for example, was developed from the zero-dynamics theory of nonlinear control \citep{byrnes1984frequency,isidori1995nonlinear} specifically for bipedal walking \citep{grizzle2001asymptotically,westervelt2003hybrid}, and its first control-community publications were already about robots. The lags range from one year (flatness) to 37 years (differential dynamic programming), with a median of about 15 years. Because the adoption dates are bounds from a targeted search and the pathways are not a representative sample, we draw no conclusion about whether lags have shortened over time.

\begin{table*}[!t]
\centering
\caption{Illustrative pathways by which concepts crossed between the control and robotics communities (C: control systems; R: robotics). Origin: earliest archival publication of the concept identified; adoption: earliest archival publication identified that applies it to the other community's systems. Rows are ordered by year of origin, as in Fig.~\ref{fig:migration}.}
\label{tab:migration}
\footnotesize
\begin{tabular}{@{}r >{\raggedright\arraybackslash}p{4.1cm} >{\raggedright\arraybackslash}p{4.3cm} >{\raggedright\arraybackslash}p{6.4cm} r@{}}
\toprule
\# & Concept & Origin (community, year) & Adoption in the other community (year) & Lag (yr)\\
\midrule
1 & Dynamic programming & C, 1952 \citep{bellman1952theory} & Minimum-time manipulator trajectories, R, 1986 \citep{shin1986dynamic} & 34\\
2 & Model-reference adaptation (MIT rule) & C, 1958 \citep{annaswamy2021historical} & Adaptive manipulator control, R, 1979 \citep{dubowsky1979application} & 21\\
3 & Kalman filtering & C, 1960 \citep{kalman1960new} & Stochastic mapping, R, 1986 \citep{smith1986representation} & 26\\
4 & Differential dynamic programming & C, 1966 \citep{mayne1966second} & Robot trajectory optimisation, R, 2003 \citep{morimoto2003minimax} & 37\\
5 & Variable-structure (sliding-mode) control & C, 1967 \citep{emelyanov1967variable} & Manipulator control, R, 1978 \citep{young1978controller} & 11\\
6 & Sequential Monte Carlo estimation & C, 1969 \citep{handschin1969monte} & Monte Carlo localisation, R, 1999 \citep{dellaert1999monte} & 30\\
7 & Dissipativity and passivity & C, 1972 \citep{willems1972dissipative} & Delay-robust bilateral teleoperation, R, 1988 \citep{anderson1988bilateral} & 16\\
8 & Model predictive control & C (industry), 1978 \citep{richalet1978model} & Mobile-robot path tracking, R, 1991 \citep{ollero1991predictive} & 13\\
9 & Learning control of robot arms & R, 1978 \citep{uchiyama1978formation} & Iterative learning control for general systems, C, 1993 \citep{moore1993iterative} & 15\\
10 & Brockett's necessary condition & C, 1983 \citep{brockett1983asymptotic} & Time-varying feedback for wheeled mobile robots, R, 1991 \citep{samson1991velocity} & 8\\
11 & Zero dynamics & C, 1984 \citep{byrnes1984frequency} & Hybrid zero dynamics of bipedal walking, R, 2001 \citep{grizzle2001asymptotically} & 17\\
12 & Artificial potential fields & R, 1985 \citep{khatib1985real} & Virtual potentials for collective motion, C, 2001 \citep{leonard2001virtual} & 16\\
13 & Differential flatness & C, 1992 \citep{fliess1992systemes} & Motion planning for vehicles with trailers, R, 1993 \citep{rouchon1993flatness} & 1\\
14 & Sum-of-squares verification & C, 2000 \citep{parrilo2000structured} & Verified feedback motion planning (LQR-trees), R, 2009 \citep{tedrake2009lqr} & 9\\
15 & Hamilton-Jacobi reachability & C, 2000 \citep{tomlin2000game} & Safe aerial-robot manoeuvres, R, 2010 \citep{gillula2010design} & 10\\
16 & Control barrier functions & C, 2007 \citep{wieland2007constructive} & Biped walking, R, 2015 \citep{hsu2015control} & 8\\
\bottomrule
\end{tabular}
\end{table*}

\subsection{The bridge scholars}

\Ncross{} scholars qualify independently for both lists (Table~\ref{tab:bridge}). They share a profile: mechanically or geometrically trained theorists who validated their ideas on hardware. Slotine took sliding and adaptive control onto manipulators. Arimoto showed that simple PD-plus-gravity control of robot arms is provably stable, and that repeated robot tasks can be learned. Brockett proved what smooth feedback cannot achieve for nonholonomic robots. Spong connected passivity to elastic joints and teleoperation. Grizzle and Ames took hybrid zero dynamics and barrier functions to walking machines and exoskeletons. Leonard took collective-motion theory to underwater gliders. Tedrake brought optimisation-based verification and open-source simulation to underactuated robots. D'Andrea and Frazzoli translated distributed control and planning into fleets of warehouse robots and autonomous vehicles. Six further scholars were scored in both fields: Murray, Sastry and Tomlin are listed in control only, and Bullo, Egerstedt and Pettersen fall outside both lists, a reminder that the lists' boundaries are porous.

\begin{table*}[!t]
\centering
\caption{The bridge scholars: individuals who qualify independently for both lists. Each field's score uses only the work relevant to that field (Section~\ref{sec:scope}).}
\label{tab:bridge}
\footnotesize
\begin{tabular}{@{}l cc cc p{9.2cm}@{}}
\toprule
& \multicolumn{2}{c}{Control} & \multicolumn{2}{c}{Robotics} & \\
\cmidrule(lr){2-3}\cmidrule(lr){4-5}
Scholar & Rank & Score & Rank & Score & Bridging contribution\\
\midrule
Jean-Jacques E. Slotine & 27 & 73.0 & 52 & 70.5 & Sliding and adaptive control proved on manipulators; contraction analysis\\
Raffaello D'Andrea & 81 & 64.5 & 16 & 77.5 & Distributed control theory turned into warehouse robot fleets\\
Russ Tedrake & 78 & 65.0 & 36 & 73.5 & Optimisation-based control and verification for underactuated robots; Drake\\
Jessy W. Grizzle & 46 & 69.5 & 70 & 68.0 & Hybrid zero dynamics: nonlinear control theory realised in bipedal robots\\
Suguru Arimoto & 60 & 67.0 & 68 & 68.0 & Passivity-based PD control and iterative learning, born in robot arms\\
Emilio Frazzoli & 87 & 64.0 & 53 & 70.5 & Maneuver automata and asymptotically optimal sampling-based planning\\
Roger W. Brockett\textdagger & 47 & 69.5 & 98 & 64.5 & Geometric control; necessary condition for smooth stabilisation of nonholonomic robots\\
Naomi Ehrich Leonard & 64 & 67.0 & 84 & 66.0 & Collective-motion control theory deployed on underwater glider fleets\\
Mark W. Spong & 72 & 66.0 & 86 & 66.0 & Elastic-joint and passivity-based robot control; delay-robust teleoperation\\
Aaron D. Ames & 70 & 66.0 & 100 & 64.0 & Control barrier functions and hybrid zero dynamics on legged robots and exoskeletons\\
\bottomrule
\end{tabular}
\end{table*}

\subsection{Communities and venues}

The two communities developed separate institutions. Control coalesced around IFAC (founded 1957), the IEEE Control Systems Society and its predecessor professional group, \emph{Automatica} (1963), the \emph{IEEE Transactions on Automatic Control} and the American and European control conferences. Robotics built its own venues: the \emph{International Journal of Robotics Research} (1982), the Robotics Society of Japan (1983), ICRA (1984), the \emph{IEEE Journal of Robotics and Automation} (1985, later the \emph{Transactions on Robotics}), IROS (1988), Robotics: Science and Systems (2005), \emph{Science Robotics} (2016) and the Conference on Robot Learning (2017). Since 2016-2018, letters journals in both societies and the \emph{Annual Review of Control, Robotics, and Autonomous Systems} have provided shared ground.

\subsection{Divergence and reconvergence}

The fields were closely coupled in the 1980s, when manipulator control was a showcase for nonlinear and adaptive theory. They diverged from the mid-1990s, as robotics turned to perception, planning, probabilistic estimation and, later, machine learning, while control turned to robustness, hybrid systems and networks. They are reconverging around a shared problem: certifying learning-enabled machines that act among people. Barrier functions, reachability, MPC-based safety filters and data-driven predictive control \citep{ames2017control,fisac2019general,hewing2020learning,brunke2022safe,garg2024advances,garg2024learning} are the common vocabulary of that reconvergence. The bridge scholars are disproportionately represented among its leaders.

\section{Future research directions}\label{sec:future}

\subsection{How the outlook was derived}

\begin{figure*}[!t]
\centering
\includegraphics[width=\textwidth]{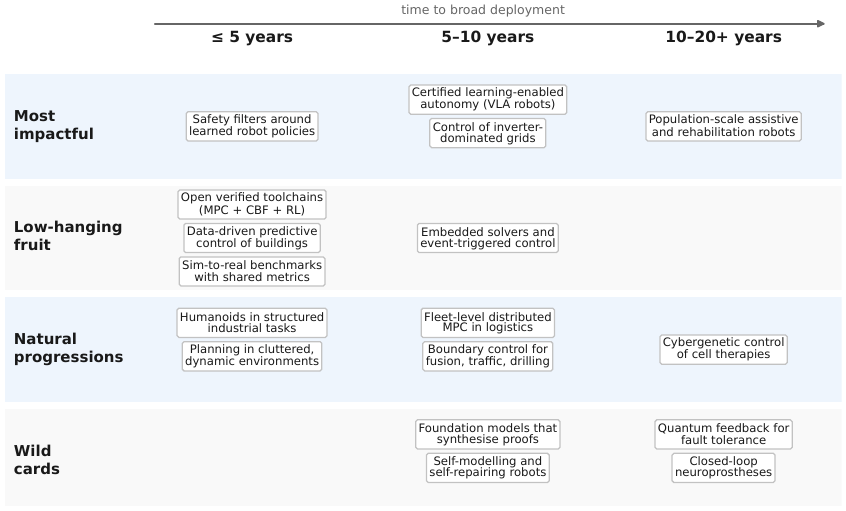}
\caption{Map of research directions by category and by expected time to broad deployment. Placement is a qualitative judgement explained in Section~\ref{sec:future} and Table~\ref{tab:future}.}
\label{fig:future}
\end{figure*}

Forecasts are the least verifiable part of any review. We derive ours from three sources. The first is the community roadmaps \citep{lamnabhi2017systems,annaswamy2023control,christensen2021roadmap,yang2018grand}. The second is the trajectories visible in the ranked careers, the concept migrations of Section~\ref{sec:crossroads} and recent results up to late 2026. The third is the pattern, evident throughout Sections~\ref{sec:control} and~\ref{sec:robotics}, that the highest impact follows when a clean abstraction meets a deployment partner. Directions are classified by expected impact and by time to broad deployment (Fig.~\ref{fig:future}). Table~\ref{tab:future} adds enablers, risks and indicators against which the forecast can later be checked.

\subsection{Most impactful}

\emph{Certified learning-enabled autonomy.} Vision-language-action models now control robots across many embodiments \citep{brohan2023rt2,openx2024,black2024pi0}, but they offer no guarantees. The highest-impact direction for both fields is to wrap such policies in runtime certificates. Safety filters built on barrier functions, reachability or predictive control can already be placed around learned policies \citep{ames2017control,fisac2019general,hewing2020learning,garg2024advances}. Certifying complete learning-enabled systems will take longer, particularly in multi-robot settings \citep{garg2024learning} and in contact-rich manipulation, where nonsmooth mechanics complicates every guarantee \citep{brogliato2023modeling}. \emph{Control of inverter-dominated power grids.} Low-inertia grids require new synchronisation and grid-forming control \citep{dorfler2013synchronization,milano2018foundations}. The stakes are decarbonisation and reliability for billions of people. \emph{Assistive and rehabilitation robotics at population scale.} Ageing societies need exoskeletons, prostheses and therapy robots whose controllers adapt to individual users \citep{lo2010robot,au2009powered,almeida2025bio}; the evidence base is growing, but scale, cost and personalisation are not yet solved.

\subsection{Low-hanging fruit}

Several advances need no new theory, only engineering consolidation. The first is open, verified toolchains that combine MPC, barrier-function filters and learned components, building on existing solvers and simulators \citep{tedrake2019drake,todorov2012mujoco}. The second is data-driven predictive control of buildings and process plants, where behavioural methods can replace costly first-principles modelling \citep{markovsky2021behavioral,berberich2021data}. The third is shared sim-to-real benchmarks with standardised metrics \citep{tobin2017domain}. The fourth, on a slightly longer horizon, is resource-aware embedded control: code-generated convex solvers and event-triggered loops running on drones and other small robots \citep{mattingley2012cvxgen,heemels2012introduction,yu2026bibliographical}.

\subsection{Natural progressions}

The established lines will continue along predictable paths. Humanoids will move into structured industrial tasks, building on learned and model-based locomotion \citep{kajita2003biped,miki2022learning}. Fleet-level distributed MPC will spread in logistics, following the warehouse model \citep{wurman2008coordinating}. Boundary control will be applied to fusion plasmas, traffic and drilling \citep{krstic2008boundary}. Sampling-based and task-and-motion planning will be refined for cluttered, dynamic environments \citep{garrett2021integrated,venu2026optimized}. Cybergenetic control of engineered cells will reach therapy \citep{aoki2019universal}.

\subsection{Wild cards}

Four low-probability directions could have outsized impact. The first is foundation models that synthesise controllers together with proofs of their correctness, closing the loop between learning and verification. The second is self-modelling and self-repairing robots \citep{bongard2006resilient}. The third is quantum feedback as an enabler of fault-tolerant quantum computing \citep{wiseman2010quantum,dong2010quantum}. The fourth is closed-loop neuroprostheses that restore voluntary movement after paralysis, building on epidural stimulation results \citep{harkema2011effect}.

\begin{table*}[!t]
\centering
\caption{Research directions with horizons, enablers, risks and indicators of progress.}
\label{tab:future}
\footnotesize
\begin{tabular}{@{}>{\raggedright\arraybackslash}p{1.9cm} >{\raggedright\arraybackslash}p{3.2cm} c >{\raggedright\arraybackslash}p{3.4cm} >{\raggedright\arraybackslash}p{2.8cm} >{\raggedright\arraybackslash}p{3.2cm}@{}}
\toprule
Category & Direction & Horizon (yr) & Enablers & Key risks & Indicators of progress\\
\midrule
Most impactful & Safety filters around learned policies & $\le$5 & Barrier functions; reachability; predictive safety filters & Conservatism; model error & Deployed robots with certified safety layers\\
 & Certified learning-enabled autonomy & 5-10 & VLA models; runtime monitors; verification & Unverifiable failures; liability & Certification standards cite runtime filters; insurer acceptance\\
 & Inverter-dominated grid control & 5-10 & Grid-forming control; synchronisation theory & Instability events; slow regulation & Share of inverter-based generation operated without stability incidents\\
 & Population-scale assistive robotics & 10-20 & Personalised control; low-cost actuation & Cost; clinical evidence & Reimbursed therapies; device fleets in homes\\
Low-hanging & Open verified toolchains & $\le$5 & Mature solvers and simulators & Fragmentation & Adoption of shared toolchains\\
 & Data-driven predictive control of buildings & $\le$5 & Behavioural theory; sensors & Data quality & Commissioned deployments; energy savings\\
 & Sim-to-real benchmarks & $\le$5 & Simulators; open datasets & Benchmark overfitting & Cross-laboratory reproducibility\\
 & Embedded solvers and event-triggered control & 5-10 & Code generation; low-power processors & Verification of embedded code & On-board optimisation in commercial drones\\
Natural & Humanoids in structured industry & $\le$5 & Learned locomotion; whole-body MPC & Safety around workers & Hours of unsupervised operation\\
 & Planning in cluttered, dynamic environments & $\le$5 & Sampling-based and task-and-motion planners & Real-time guarantees & Success rates on shared benchmarks\\
 & Fleet-level distributed MPC & 5-10 & Cloud-edge compute & Cyber-security & Fleet size under coordinated control\\
 & Boundary control of PDEs & 5-10 & Backstepping; sensing & Model mismatch & Field trials in fusion, traffic, drilling\\
 & Cybergenetic cell therapies & 10-20 & Synthetic biology & Biological variability & Clinical-stage controllers\\
Wild card & Proof-producing foundation models & 5-10 & Formal methods; large models & Spurious proofs & Verified controllers generated automatically\\
 & Self-modelling, self-repairing robots & 5-10 & Learned self-models; modular hardware & Reliability & Autonomous recovery from damage\\
 & Quantum feedback & 10-20 & Quantum hardware & Decoherence & Feedback-stabilised logical qubits\\
 & Closed-loop neuroprostheses & 10-20 & Neural interfaces & Safety; ethics & Restored function in trials\\
\bottomrule
\end{tabular}
\end{table*}

\section{How far can the lists be trusted?}\label{sec:robustness}

\subsection{Alternative weightings}

Table~\ref{tab:sensitivity} reports agreement between the baseline and the four alternative schemes, each ranked with the tie rule of Section~\ref{sec:ties}. Spearman correlations lie between \rhoMin{} and \rhoMax{} and Kendall correlations between \tauMin{} and \tauMax{}. The impact-heavy and no-recognition schemes retain \CovlImp{} and \CovlNoR{} of the control list and \RovlImp{} and \RovlNoR{} of the robotics list. The citation-heavy scheme is the most disruptive (\CovlCit{} and \RovlCit{} retained). It promotes highly cited theorists and demotes deployment-oriented scholars such as Makino, Mann and Krebs, which is precisely the effect an impact-first rubric is designed to avoid.

\begin{table*}[!t]
\centering
\caption{Rank stability. Upper block: alternative weighting schemes, each ranked with the same tie rule; $\rho$: Spearman and $\tau$: Kendall rank correlation with the baseline over the full scored pool; overlap: baseline top-100 (top-10) scholars retained; $|\Delta r|$: median and maximum rank shift of baseline top-100 scholars. Lower block: number of listed scholars whose inclusion frequency is at least 0.95 under each perturbation model, and number of candidates in the whole pool whose frequency lies strictly between 0.05 and 0.95.}
\label{tab:sensitivity}
\footnotesize
\setlength{\tabcolsep}{4pt}
\begin{tabular}{@{}l l cccccc cc c c@{}}
\toprule
Field & Scheme & $w_I$ & $w_F$ & $w_S$ & $w_T$ & $w_E$ & $w_R$ & $\rho$ & $\tau$ & Top-100 / Top-10 & $|\Delta r|$ med./max\\
\midrule
Control & Baseline (impact-first) & 0.35 & 0.20 & 0.15 & 0.10 & 0.10 & 0.10 & 1.000 & 1.000 & 100 / 10 & 0 / 0\\
 & Equal weights & 0.17 & 0.17 & 0.17 & 0.17 & 0.17 & 0.17 & 0.985 & 0.913 & 96 / 7 & 3 / 46\\
 & Impact-heavy (I = 50\%) & 0.50 & 0.15 & 0.12 & 0.08 & 0.08 & 0.08 & 0.989 & 0.929 & 96 / 10 & 4 / 33\\
 & Citation-heavy (S = 35\%) & 0.20 & 0.15 & 0.35 & 0.10 & 0.10 & 0.10 & 0.962 & 0.843 & 93 / 8 & 10 / 70\\
 & No peer recognition (R = 0) & 0.39 & 0.22 & 0.17 & 0.11 & 0.11 & 0.00 & 0.992 & 0.932 & 98 / 9 & 4 / 17\\
\midrule
Robotics & Baseline (impact-first) & 0.35 & 0.20 & 0.15 & 0.10 & 0.10 & 0.10 & 1.000 & 1.000 & 102 / 10 & 0 / 0\\
 & Equal weights & 0.17 & 0.17 & 0.17 & 0.17 & 0.17 & 0.17 & 0.985 & 0.915 & 97 / 9 & 5 / 31\\
 & Impact-heavy (I = 50\%) & 0.50 & 0.15 & 0.12 & 0.08 & 0.08 & 0.08 & 0.992 & 0.950 & 98 / 9 & 3 / 18\\
 & Citation-heavy (S = 35\%) & 0.20 & 0.15 & 0.35 & 0.10 & 0.10 & 0.10 & 0.941 & 0.811 & 93 / 8 & 10 / 75\\
 & No peer recognition (R = 0) & 0.39 & 0.22 & 0.17 & 0.11 & 0.11 & 0.00 & 0.994 & 0.948 & 99 / 9 & 3 / 16\\
\midrule
\multicolumn{2}{@{}l}{Perturbation model} & \multicolumn{6}{l}{Draws} & \multicolumn{2}{c}{Control} & \multicolumn{2}{c}{Robotics}\\
& & \multicolumn{6}{l}{} & $\geq0.95$ & marginal & $\geq0.95$ & marginal\\
\cmidrule(lr){9-10}\cmidrule(l){11-12}
\multicolumn{2}{@{}l}{Weights, $W$ (impact-first Dirichlet)} & \multicolumn{6}{l}{20{,}000} & 81 & 34 & 87 & 28\\
\multicolumn{2}{@{}l}{Scores, $S$ ($\pm1$ per criterion)} & \multicolumn{6}{l}{20{,}000} & 58 & 86 & 60 & 84\\
\multicolumn{2}{@{}l}{Joint, $J$ (weights and scores)} & \multicolumn{6}{l}{20{,}000} & 54 & 92 & 55 & 90\\
\bottomrule
\end{tabular}
\end{table*}

\subsection{Perturbed weights and perturbed scores}

\begin{figure*}[!t]
\centering
\includegraphics[width=\textwidth]{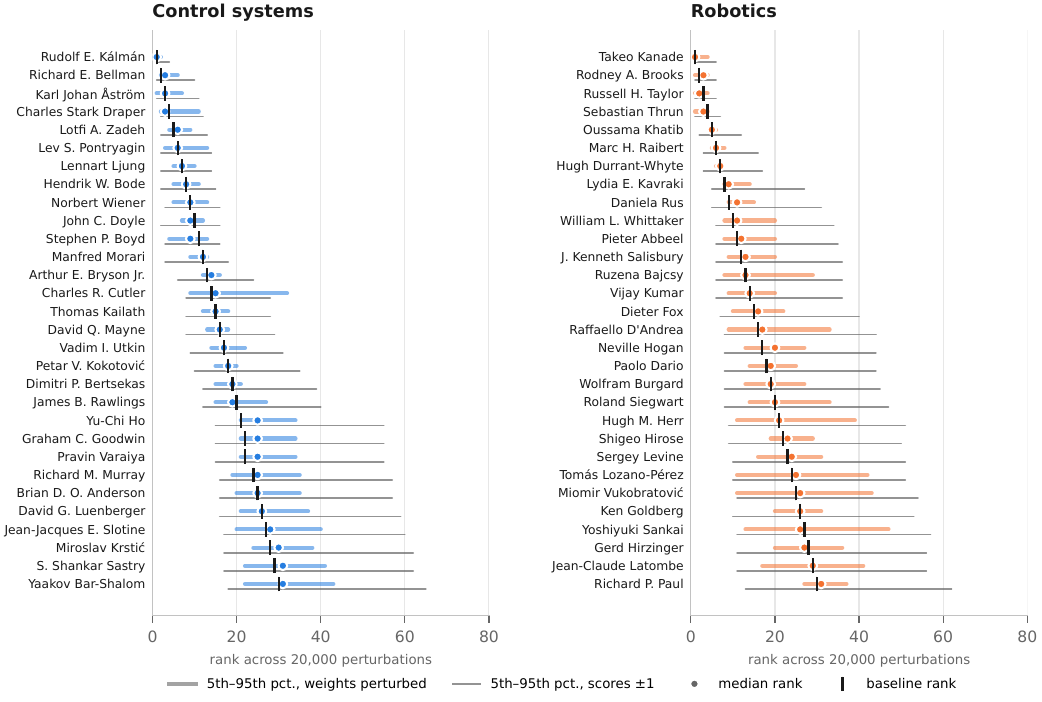}
\caption{Rank robustness of the thirty highest-ranked scholars in each field. Thick lines: 5th-95th percentiles of rank under \NCmc{} Dirichlet-distributed weight vectors that keep impact on humanity as the largest weight (model W); thin grey lines: the same under \Nsc{} independent $\pm1$ perturbations of every criterion score (model S); dots: median rank under W; ticks: baseline rank.}
\label{fig:robustness}
\end{figure*}

Fig.~\ref{fig:robustness} shows the rank distributions of the thirty highest-ranked scholars in each field under the weight model W and the score model S. Under W, \NCmcsecure{} control and \NRmcsecure{} robotics scholars in the lists have an inclusion frequency of at least 0.95; in control, \NCtopten{} of the baseline top ten have a median rank within the top ten, and in robotics \NRtopten{} do. Perturbing the scores is more disruptive than perturbing the weights: under S, \NCscsecure{} control and \NRscsecure{} robotics scholars reach 0.95, and under the joint model J, \NCjtsecure{} and \NRjtsecure{} (Table~\ref{tab:sensitivity}). The leading places are secure under all three models; the places in the lower third of each list are not. Near the cut-off, membership is a matter of near-ties rather than clear separation: the five candidates on either side of rank 100 have $W_{100}$ between \CbandLo{} and \CbandHi{} in control and between \RbandLo{} and \RbandHi{} in robotics, and $S_{100}$ between \CsbandLo{} and \CsbandHi{} and between \RsbandLo{} and \RsbandHi{} (Table~\ref{tab:hm}). Ames, Ishiguro and Sitti, whose identical vectors share rank 100 in robotics, now receive identical frequencies under every model, as they must. The rank intervals are widest for scholars with uneven profiles: theorists whose scholarly influence far exceeds their direct technology transfer (Isidori and Sontag in control; Lozano-Pérez and Vukobratović in robotics), and builders with the reverse profile (Cutler, Sankai). We read these results as a statement about precision: the lists separate the leading scholars of each field clearly, but the order within the lower half, and membership within roughly fifteen places of the cut-off, should not be over-interpreted.

\subsection{Consistency with award rosters}\label{sec:concordance}

Table~\ref{tab:awards} compares the lists with the six identification rosters. Of the \Nrosterscored{} scored roster recipients (counted once per roster), \Nrosterin{} are listed. The control list contains \NCSAin{} of the \NCSA{} IEEE Control Systems Award recipients, \NQUAin{} of \NQUA{} Quazza medallists, \NBODin{} of \NBOD{} Bode lecturers and \NBELin{} of \NBEL{} scored Bellman Heritage awardees; the robotics list contains \NRAAin{} of \NRAA{} scored IEEE Robotics and Automation Award recipients and \NPIOin{} of \NPIO{} scored RAS Pioneer awardees. The Bellman award, which honours lifetime contributions to control practice and education, has the lowest share, because many of its earlier recipients made contributions whose documented impact on deployed technology is more modest than that of the scholars listed. The IEEE Control Systems Award recipients outside the control list (Morse, Vidyasagar, Krasovskii, Mitter, Youla, Bensoussan, Zakai, Lindquist and Kimura) are distinguished theorists; their exclusion is the intended consequence of weighting impact first, and Morse and Vidyasagar fall just outside, at a shared rank of 104. This agreement is a consistency check, not independent validation: the same rosters were used to identify candidates and contribute to the recognition criterion. Setting $w_R=0$ changes the counts little (Table~\ref{tab:awards}), which shows that the agreement is not driven by the recognition score itself; it does not remove the selection link, since every roster recipient was guaranteed a place in the scored pool.

\begin{table*}[!t]
\centering
\caption{Consistency check against the six identification rosters. Because the same rosters were used to identify candidates and enter the recognition criterion, agreement is partly built in and is not independent validation (Section~\ref{sec:concordance}). Recipients: distinct individuals on the roster; screened: recipients without a qualifying academic appointment or outside the field (codes A/B, Appendix~\ref{app:protocol}); share: listed recipients among those scored; $w_R=0$: listed recipients when recognition is removed from the score.}
\label{tab:awards}
\footnotesize
\begin{tabular}{@{}>{\raggedright\arraybackslash}p{5.6cm} l c c c c c c@{}}
\toprule
Roster (years) & Field & Recipients & Screened & Scored & Listed & Share & $w_R=0$\\
\midrule
IEEE Control Systems Award (1982--2027) & Control & 45 & 0 & 45 & 36 & 80\% & 34\\
AACC Bellman Control Heritage Award (1979--2026) & Control & 48 & 4 & 44 & 28 & 64\% & 27\\
IFAC Quazza Medal (1981--2023) & Control & 15 & 0 & 15 & 12 & 80\% & 12\\
IEEE CSS Bode Lecture Prize (1989--2025) & Control & 37 & 0 & 37 & 26 & 70\% & 25\\
IEEE Robotics and Automation Award (2004--2027) & Robotics & 28 & 5 & 23 & 21 & 91\% & 20\\
IEEE RAS Pioneer Award (1999--2026) & Robotics & 44 & 2 & 42 & 36 & 86\% & 34\\
\bottomrule
\end{tabular}
\end{table*}

\subsection{Composition of the lists}\label{sec:composition}

\begin{figure*}[!t]
\centering
\includegraphics[width=\textwidth]{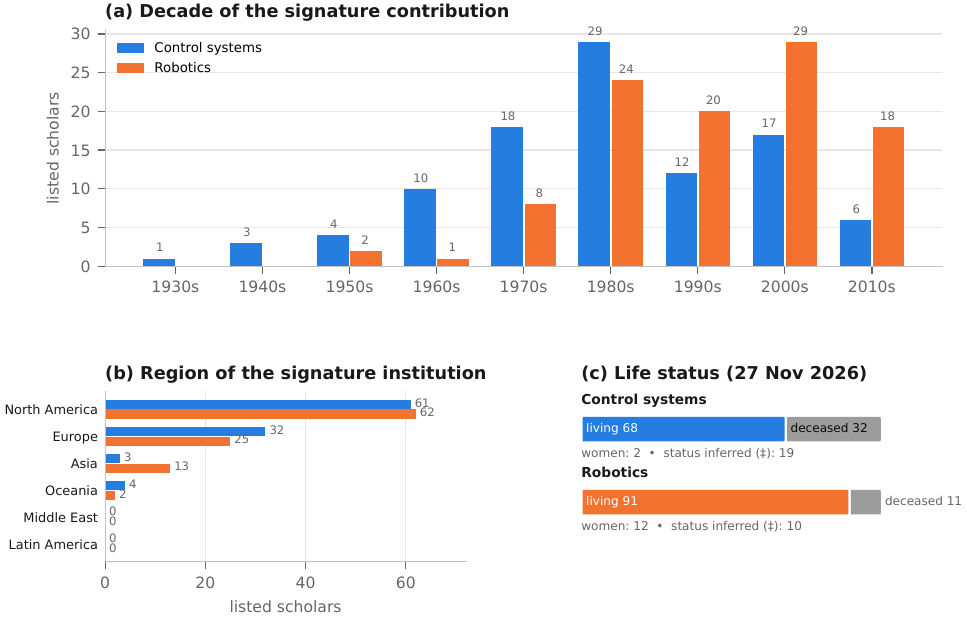}
\caption{Composition of the two lists: (a) decade of the signature contribution; (b) region of the institution at which the signature work was done; (c) life status on 27 November 2026, number of women, and number of living scholars whose status is inferred (\ddag).}
\label{fig:composition}
\end{figure*}

Fig.~\ref{fig:composition} summarises the composition of the two lists.

\paragraph{Era} The signature contributions in control peak in the \CeraPeak s (\NCeraPeak{} scholars) and in robotics in the \ReraPeak s (\NReraPeak{}). This reflects robotics' later maturity and its rapid growth since 2000.

\paragraph{Geography} North American institutions host \NCregNA{} control and \NRregNA{} robotics signature contributions, Europe \NCregEU{} and \NRregEU{}, Asia \NCregAS{} and \NRregAS{}, and Oceania \NCregOC{} and \NRregOC{}. No listed signature contribution was made at an institution in the Middle East, Latin America or Africa, although several listed scholars were born or trained there. Asia's strong presence in robotics (Japan above all) contrasts with its near-absence in control. In part this reflects the language barrier, and in part the recency of China's rapid growth in control research, whose impact may become visible in future assessments.

\paragraph{Gender} The lists include \NCwomen{} women in control and \NRwomen{} in robotics, a historical under-representation that the impact criterion cannot correct and which mirrors who was admitted to and recognised in these fields \citep{rossiter1993matthew}.

\paragraph{Life status} The control list includes \NCdead{} deceased scholars and the robotics list \NRdead{}. Seven scholars in the control list have died since 2022 (Utkin, Brockett, Wonham, Mayne and, in 2026, Luenberger, Stein and Åström), and four in the robotics list (Brockett, Bekey, Paul and Mori). Their passing marks the end of the generation that built modern control theory and the first robot laboratories.

\paragraph{Nature of contribution} Theory appears in the contribution profile of \NCnatTh{} of the \NCtop{} control scholars, against \NRnatTh{} of the \NRtop{} robotics entries. In robotics, hardware and systems contributions dominate (\NRnatHw{} against \NCnatHw{} in control). Education is a defining element for \NCnatEd{} control and \NRnatEd{} robotics entries.

\section{Discussion}\label{sec:discussion}

\subsection{What the highest-ranked careers have in common}

Across both lists, the highest impact scores almost always combine three elements: a clean mathematical or architectural abstraction; a partner that deployed it at scale; and a teachable artefact, such as a textbook, toolbox or open-source package, that propagated it. Kálmán's filter met NASA Ames \citep{mcgee1985discovery}; Åström's adaptive and auto-tuning ideas met Swedish process industry; Boyd's convex formulations met embedded code generation \citep{mattingley2012cvxgen}; Durrant-Whyte's estimation methods met container-port operators \citep{durrantwhyte2007autonomous}; Brooks's architecture met a consumer product \citep{jones2006robots}; Taylor's surgical robot met surgeons \citep{taylor1994image}; Thrun's probabilistic methods met the DARPA challenges \citep{thrun2006stanley}. For early-career researchers, the implication is that impact is designed, not awaited: abstraction, partnership and pedagogy must be pursued together. Mentoring compounds all three, as the lineages of Fig.~\ref{fig:genealogy} show.

\subsection{Ethics and societal responsibility}

The impact criterion counts benefits, but the same technologies carry risks. Autonomy in weapons raises unresolved questions of accountability \citep{arkin2009governing,lin2012robot}. Industrial robots have measurable effects on employment and wages \citep{acemoglu2020robots}. Increasing autonomy in medical robots raises regulatory, ethical and legal questions \citep{yang2017medical}. The concentration of data and computation in a few firms shapes who benefits from learning-enabled control. A ranking of people also carries an ethical responsibility. It risks reinforcing hierarchies and the visibility of those already visible, so we have published the anchors, scores and screened names, credited industrial originators and teams, and avoided ranking early-career scholars.

\subsection{Limitations}

The principal limitation is that the scores are structured judgements. The revision makes them auditable, through a source-backed record for every candidate, and tests them, through score perturbations; independent reassessment of the candidates near the cut-off is the remaining safeguard (Section~\ref{sec:reassess}). Second, identification is complete only with respect to the six award rosters; beyond them it rests on a targeted search, and the inclusive eligibility rule, while now applied uniformly, is itself a modelling choice. Third, impact is hard to attribute, and our evidence for it is uneven: strong for aerospace, process control and surgery, weaker for theory whose influence is diffuse. Fourth, coverage of non-English scholarship, and of scholars outside North America, Europe and East Asia, is incomplete. Fifth, life status for the oldest living scholars is inferred, not confirmed. Finally, the field boundaries, especially at the borders with AI, optimisation and networking, exclude some influential people. We regard the lists as a reasoned, contestable baseline, and the released evidence, data and code as an invitation to improve it.

\section{Conclusions}\label{sec:conclusions}

Reading the history of control and robotics through its people shows how ideas reached practice. The highest-impact careers paired a clean abstraction with a deployment partner and a teachable artefact; schools and lineages compounded those advantages; and concepts repeatedly crossed between the two communities, often carried by scholars who worked on both sides. In the sixteen well-documented pathways we examined, most crossings ran from control to robotics, but learning control and potential fields travelled the other way and some ideas, such as hybrid zero dynamics, were born at the crossroads. The two fields are now reconverging around a shared problem, certifying learning-enabled machines that act among people, which heads our research agenda together with control of inverter-dominated grids and assistive robotics at population scale.

Weighting impact on humanity first yields a control list led by the founders of estimation, optimal control and adaptive control, the pioneers of inertial guidance, frequency-domain design and fuzzy logic, and the originator of industrial predictive control, and a robotics list led by builders whose work reached roads, homes, operating rooms, factories and warehouses. The leading places are stable under perturbed weights and scores, whereas membership near the cut-off is fragile and is the subject of independent reassessment. The lists are consistent with peer-selected award rosters, a check rather than a validation. Every score, exclusion and figure can be traced to released evidence and code, so that readers can contest the judgements as well as re-weight them.

\section*{CRediT authorship contribution statement}
\textbf{Ali Nasir:} Conceptualization, Methodology, Investigation, Data curation, Formal analysis, Software, Visualization, Validation, Writing - original draft, Writing - review \& editing.

\section*{Declaration of competing interest}
The author declares that there are no known competing financial interests or personal relationships that could have appeared to influence the work reported in this paper.

\section*{Declaration of generative AI and AI-assisted technologies in the manuscript preparation process}
During the preparation of this work the author used Claude (Anthropic) Opus 5.5 model in order to draft and edit the text and tables of the manuscript and the supplementary scoring record. The use of the same tool in the research process (source retrieval, initial scoring proposals, and the code that computes the rankings and renders the figures) is described in Section~\ref{sec:aiuse}. No generative image model was used; all figures are rendered by the released code from the released data. After using this tool, the author reviewed and edited the content as needed and takes full responsibility for the content of the published article.

\section*{Data availability}
The criterion scores, evidence and life status (confirmed or inferred) of all \Npooltotal{} scored candidates (Supplementary Material S1 and \texttt{scores\_*.csv}), the screening log with exclusion codes and sources, the award rosters, the blind sheets and analysis script for independent reassessment, and the code that generates every ranking, table and figure are provided as supplementary material (folder \texttt{supplementary/}) and will be deposited in a public repository upon acceptance.

\section*{Funding}
This research did not receive any specific grant from funding agencies in the public, commercial, or not-for-profit sectors.

\appendix
\renewcommand{\theHtable}{app.\Alph{section}.\arabic{table}}

\section{The ranked lists}\label{app:lists}
\setcounter{table}{0}

Tables~\ref{tab:control100} and~\ref{tab:robotics100} give the complete lists. Asterisks mark candidates added in the revised screening (Section~\ref{sec:screening}).

\startwide
\begin{landscape}
\setlength{\LTcapwidth}{\linewidth}
\scriptsize
\setlength{\tabcolsep}{2.6pt}
\renewcommand{\arraystretch}{1.04}
\begin{longtable}{@{}r >{\raggedright\arraybackslash}p{3.15cm} >{\raggedright\arraybackslash}p{3.2cm} >{\raggedright\arraybackslash}p{6.6cm} >{\raggedright\arraybackslash}p{1.4cm} >{\raggedright\arraybackslash}p{4.6cm} cccccc r@{}}
\caption{The top 100 scholars in control systems (1926--2026), ranked by the impact-first composite score.}\label{tab:control100}\\
\toprule
\# & Scholar & Key institution(s) & Signature contribution(s) & Nature & Impact on humanity & I & F & S & T & E & R & Score\\
\midrule
\endfirsthead
\multicolumn{13}{@{}l}{\textit{Table \thetable\ (continued)}}\\
\toprule
\# & Scholar & Key institution(s) & Signature contribution(s) & Nature & Impact on humanity & I & F & S & T & E & R & Score\\
\midrule
\endhead
\midrule
\multicolumn{13}{r@{}}{\textit{continued on next page}}\\
\endfoot
\bottomrule
\multicolumn{13}{@{}p{23.85cm}@{}}{\textdagger\ deceased (life span under the name); \ddag\ living status inferred from the absence of any death report as of 27~November~2026 (scholar born in or before 1945, or birth year unverified); [X] in both lists; * added in the revised screening. Nature: Th theory, Al algorithm, Hw hardware/systems, Sw software, Ed education. Criteria (0--10): I impact on humanity (weight 35\%), F field creation (20\%), S scholarly influence (15\%), T technology transfer (10\%), E education and community (10\%), R peer recognition (10\%). Score $=3.5I+2F+1.5S+T+E+R$. Equal scores are ordered by I, then F, S, T, E and R; identical criterion vectors share a rank and are listed alphabetically. The evidence behind every score is in Supplementary Material S1.}\\
\endlastfoot

1 & \textbf{Rudolf E. K\'alm\'an}\textdagger\newline{\tiny 1930--2016} & Stanford; Univ. of Florida; ETH Z\"urich & Kalman filter; state-space realisation; controllability and observability \citep{kalman1960new,kalman1961new} & Th/Al & Estimation core of Apollo navigation, aviation and GNSS receivers \citep{mcgee1985discovery,grewal2010applications} & 10 & 10 & 10 & 7 & 6 & 10 & 93.0\\
2 & \textbf{Richard E. Bellman}\textdagger\newline{\tiny 1920--1984} & RAND; Univ. of Southern California & Dynamic programming; principle of optimality; Hamilton--Jacobi--Bellman equation \citep{bellman1957dynamic} & Th/Al & Basis of sequential decision making, from logistics to reinforcement learning & 9 & 10 & 10 & 6 & 7 & 9 & 88.5\\
3 & \textbf{Karl Johan \AA str\"om}\textdagger\newline{\tiny 1934--2026} & Lund University & Self-tuning regulators; relay auto-tuning of PID; stochastic control; event-based sampling \citep{astrom1973selftuning,astrom1984automatic,astrom2002comparison} & Th/Al/Ed & PID auto-tuners and adaptive controllers in process industry & 8 & 8 & 10 & 9 & 10 & 10 & 88.0\\
4 & \textbf{Charles Stark Draper}\textdagger\newline{\tiny 1901--1987} & MIT (Instrumentation Laboratory) & Inertial guidance and navigation; Apollo guidance, navigation and control \citep{draper1960inertial,hall1996journey} & Hw/Ed & Inertial navigation of aircraft, ships and spacecraft; Apollo landings & 10 & 8 & 6 & 10 & 8 & 9 & 87.0\\
5 & \textbf{Lotfi A. Zadeh}\textdagger\newline{\tiny 1921--2017} & Columbia; UC Berkeley & Fuzzy sets and fuzzy logic; state-space linear system theory \citep{zadeh1965fuzzy,zadeh1963linear} & Th & Fuzzy control in consumer products, transport and industry & 8 & 10 & 10 & 7 & 7 & 9 & 86.0\\
6 & \textbf{Lev S. Pontryagin}\textdagger\newline{\tiny 1908--1988} & Moscow State Univ.; Steklov Institute & Maximum principle of optimal control \citep{pontryagin1962mathematical} & Th & Optimal trajectories for launch, orbital transfer and guidance & 9 & 10 & 9 & 4 & 7 & 9 & 85.0\\
7 & \textbf{Lennart Ljung} & Link\"oping University & Prediction-error framework for system identification; System Identification Toolbox \citep{ljung1999system} & Th/Sw & Data-driven modelling of engines, aircraft and process plants & 8 & 8 & 10 & 9 & 8 & 9 & 85.0\\
8 & \textbf{Hendrik W. Bode}\textdagger\newline{\tiny 1905--1982} & Bell Labs; Harvard University & Frequency-domain feedback design; gain--phase relations; sensitivity integral \citep{bode1945network} & Th & Feedback amplifiers for long-distance telephony; servo design practice & 9 & 9 & 8 & 8 & 7 & 8 & 84.5\\
9 & \textbf{Norbert Wiener}\textdagger\newline{\tiny 1894--1964} & MIT & Cybernetics; optimal (Wiener) filtering \citep{wiener1948cybernetics,wiener1949extrapolation} & Th & Optimal filtering and a cybernetic view of feedback across disciplines & 8 & 10 & 10 & 5 & 7 & 9 & 84.0\\
10 & \textbf{John C. Doyle} & Caltech & Structured singular value ($\mu$); state-space $\mathcal{H}_2/\mathcal{H}_\infty$ synthesis \citep{doyle1982analysis,doyle1989state} & Th/Al & Robust flight-control and process-control design & 8 & 9 & 10 & 6 & 8 & 9 & 84.0\\
11 & \textbf{Stephen P. Boyd} & Stanford University & LMIs and convex optimisation in control; CVX/CVXPY; code generation for embedded solvers \citep{boyd1994linear,mattingley2012cvxgen} & Al/Sw/Ed & Embedded convex optimisation, including on-board rocket-landing guidance code \citep{blackmore2016autonomous} & 8 & 7 & 10 & 10 & 9 & 8 & 84.0\\
12 & \textbf{Manfred Morari} & Caltech; ETH Z\"urich; Univ. of Pennsylvania & Internal model control; robust, hybrid and explicit MPC \citep{garcia1982internal,morari1989robust,bemporad2002explicit} & Al/Th/Ed & Model predictive control in refining, chemicals and automotive systems \citep{qin2003survey} & 8 & 8 & 9 & 8 & 8 & 9 & 82.5\\
13 & \textbf{Arthur E. Bryson Jr.}\ddag & Harvard; Stanford & Applied optimal control and estimation; gradient methods for trajectory optimisation \citep{bryson1969applied} & Al/Ed & Trajectory optimisation and guidance of aircraft and spacecraft & 8 & 8 & 8 & 6 & 9 & 9 & 80.0\\
14 & \textbf{Charles R. Cutler}\textdagger*\newline{\tiny 1936--2020} & Shell Oil; Univ. of Houston (distinguished adjunct prof.) & Dynamic matrix control, the first widely deployed industrial MPC \citep{cutler1980dynamic} & Al/Sw & Predictive control of refineries and petrochemical plants worldwide \citep{qin2003survey} & 9 & 8 & 7 & 10 & 4 & 7 & 79.0\\
15 & \textbf{Thomas Kailath}\ddag & Stanford University & Linear systems; fast and structured estimation algorithms \citep{kailath1980linear} & Th/Al/Ed & Estimation algorithms in communications and signal processing & 7 & 7 & 9 & 8 & 9 & 10 & 79.0\\
16 & \textbf{David Q. Mayne}\textdagger\newline{\tiny 1930--2024} & Imperial College London & Stability theory of MPC; differential dynamic programming; Monte Carlo filtering \citep{mayne2000constrained,jacobson1970differential,handschin1969monte} & Th/Al & Theoretical guarantees behind deployed predictive control & 8 & 8 & 9 & 5 & 7 & 9 & 78.5\\
17 & \textbf{Vadim I. Utkin}\textdagger\newline{\tiny 1937--2022} & Inst. of Control Sciences (Moscow); Ohio State Univ. & Sliding-mode control, with Emelyanov; equivalent control \citep{utkin1977variable,utkin1992sliding} & Th/Al & Robust control of electric drives, converters and vehicles & 8 & 9 & 8 & 6 & 7 & 7 & 78.0\\
18 & \textbf{Petar V. Kokotovi\'c}\ddag & UIUC; UC Santa Barbara & Singular perturbations; backstepping; adaptive nonlinear control \citep{kokotovic1986singular,krstic1995nonlinear} & Th/Al/Ed & Multi-time-scale and nonlinear control of engines, power systems and aircraft & 7 & 8 & 9 & 5 & 9 & 9 & 77.0\\
19 & \textbf{Dimitri P. Bertsekas}\ddag & MIT; Arizona State Univ. & Dynamic programming and optimal control; neuro-dynamic programming \citep{bertsekas1996neuro,bertsekas2019reinforcement} & Th/Al/Ed & Algorithms for networks, logistics and reinforcement learning & 7 & 7 & 10 & 5 & 9 & 9 & 76.5\\
20 & \textbf{James B. Rawlings} & Univ. of Wisconsin--Madison; UC Santa Barbara & MPC theory and moving-horizon estimation \citep{rawlings2017model} & Th/Ed & Industrial predictive control and estimation & 8 & 6 & 8 & 8 & 8 & 8 & 76.0\\
21 & \textbf{Yu-Chi Ho}\ddag & Harvard University & Applied optimal control; differential games; perturbation analysis of discrete-event systems \citep{bryson1969applied,ho1991perturbation} & Th/Al & Optimisation of manufacturing and communication networks & 6 & 8 & 9 & 5 & 9 & 9 & 73.5\\
22 & \textbf{Graham C. Goodwin}\ddag & Univ. of Newcastle (Australia) & Adaptive and sampled-data control; fundamental limitations \citep{goodwin1980discrete,goodwin1984adaptive} & Th/Ed & Industrial control in mining, metals and telecommunications & 6 & 7 & 9 & 7 & 9 & 9 & 73.5\\
22 & \textbf{Pravin Varaiya}\ddag & UC Berkeley & Stochastic and hybrid systems; traffic and power-system control \citep{kumar1986stochastic,chen2001freeway} & Th/Sw & Freeway performance monitoring and traffic sensing & 6 & 7 & 9 & 7 & 9 & 9 & 73.5\\
24 & \textbf{Richard M. Murray} & Caltech & Nonlinear and flight control; differential flatness; consensus; biomolecular feedback \citep{murray1994mathematical,olfati2004consensus,astrom2008feedback} & Th/Hw/Ed & Autonomous vehicles and engineered biological circuits & 6 & 7 & 9 & 6 & 10 & 9 & 73.5\\
25 & \textbf{Brian D. O. Anderson}\ddag & Australian National University & Optimal filtering and linear-quadratic control; network synthesis \citep{anderson1979optimal,anderson1990optimal} & Th/Ed & Estimation and control practice; national ICT research leadership & 6 & 7 & 9 & 6 & 9 & 10 & 73.5\\
26 & \textbf{David G. Luenberger}\textdagger\newline{\tiny 1937--2026} & Stanford University & State observers; optimisation theory \citep{luenberger1964observing,luenberger1971introduction} & Th/Ed & State estimation with fewer sensors in machines and vehicles & 7 & 7 & 9 & 5 & 9 & 7 & 73.0\\
27 & \textbf{Jean-Jacques E. Slotine} [X] & MIT & Sliding and adaptive control of manipulators; contraction analysis \citep{slotine1987adaptive,lohmiller1998contraction,slotine1991applied} & Th/Al/Ed & Robust control of robots and vehicles & 6 & 8 & 10 & 5 & 9 & 7 & 73.0\\
28 & \textbf{Miroslav Krsti\'c} & UC San Diego & Backstepping for PDEs; extremum seeking; adaptive nonlinear control \citep{krstic2000stability,krstic2008boundary} & Th/Al & Extremum seeking and boundary control in engines, energy and drilling & 6 & 8 & 9 & 5 & 8 & 9 & 72.5\\
29 & \textbf{S. Shankar Sastry} & UC Berkeley & Nonlinear, adaptive and hybrid control; geometric robotics; cyber-physical security \citep{sastry1989adaptive,sastry1999nonlinear,murray1994mathematical} & Th/Ed & Theory and training behind autonomous and cyber-physical systems & 6 & 7 & 9 & 5 & 10 & 9 & 72.5\\
30 & \textbf{Yaakov Bar-Shalom}\ddag & University of Connecticut & Multitarget tracking; data association; interacting multiple model estimator \citep{barshalom1988tracking,blom1988interacting} & Th/Al & Tracking in radar, air-traffic control and automotive sensing & 7 & 7 & 9 & 6 & 7 & 7 & 72.0\\
31 & \textbf{Alberto Isidori}\ddag & Sapienza Univ. of Rome & Geometric nonlinear control; feedback linearisation; output regulation \citep{isidori1981nonlinear,isidori1990output,isidori1995nonlinear} & Th & Nonlinear design for aircraft, robots and power electronics & 6 & 8 & 10 & 3 & 8 & 9 & 72.0\\
31 & \textbf{Eduardo D. Sontag} & Rutgers; Northeastern & Input-to-state stability; control Lyapunov functions; systems biology \citep{sontag1989smooth,sontag1989universal,sontag1998mathematical} & Th & Robustness and safety certificates; models of cell signalling & 6 & 8 & 10 & 3 & 8 & 9 & 72.0\\
33 & \textbf{Gordon S. Brown}\textdagger\newline{\tiny 1907--1996} & MIT (Servomechanisms Laboratory) & Servomechanism theory; numerically controlled machine tools \citep{brown1948principles} & Hw/Ed & Origins of numerically controlled manufacturing & 7 & 7 & 6 & 8 & 9 & 7 & 71.5\\
34 & \textbf{Masayoshi Tomizuka} & UC Berkeley & Zero-phase-error tracking; repetitive and preview control; mechatronics \citep{tomizuka1987zero,tomizuka1989analysis} & Th/Al/Ed & Precision servo control in disk drives and motion systems & 7 & 6 & 8 & 6 & 9 & 8 & 71.5\\
35 & \textbf{Keith Glover} & University of Cambridge & Hankel-norm reduction; $\mathcal{H}_\infty$ synthesis; loop-shaping \citep{glover1984all,mcfarlane1992loop} & Th/Al & Robust flight-control design by loop shaping & 6 & 8 & 9 & 5 & 7 & 9 & 71.5\\
36 & \textbf{Kumpati S. Narendra}\ddag & Harvard; Yale & Stable adaptive control; neural identification and control \citep{narendra1989stable,narendra1990identification} & Th/Ed & Adaptive and learning control in practice & 6 & 8 & 9 & 4 & 9 & 8 & 71.5\\
37 & \textbf{Malcolm C. Smith} & University of Cambridge & The inerter; passive mechanical network synthesis \citep{smith2002synthesis} & Hw/Th & Inerters in racing-car and railway suspensions & 6 & 8 & 8 & 7 & 6 & 9 & 71.0\\
38 & \textbf{Stanislav V. Emelyanov}\textdagger\newline{\tiny 1929--2018} & Moscow State University & Variable-structure control systems; co-founder of sliding-mode control \citep{emelyanov1967variable} & Th & Robust control of uncertain plants & 7 & 9 & 6 & 5 & 7 & 7 & 70.5\\
39 & \textbf{Alberto Bemporad} & IMT School for Advanced Studies Lucca & Hybrid and explicit MPC; embedded QP solvers \citep{bemporad1999control,bemporad2002explicit} & Al/Sw & Embedded predictive control in vehicles & 6 & 7 & 9 & 8 & 7 & 7 & 70.5\\
40 & \textbf{Rolf Isermann}\ddag & TU Darmstadt & Model-based fault detection and diagnosis; mechatronic and automotive control \citep{isermann2005model,isermann2006fault} & Th/Al & Fault diagnosis in vehicles and industrial processes & 6 & 7 & 9 & 7 & 8 & 7 & 70.5\\
41 & \textbf{Howard H. Rosenbrock}\textdagger\newline{\tiny 1920--2010} & UMIST & Multivariable frequency-domain design; computer-aided control design \citep{rosenbrock1970state,rosenbrock1974computer} & Th/Al & Multivariable control of process plants & 6 & 8 & 8 & 5 & 7 & 9 & 70.0\\
42 & \textbf{W. Murray Wonham}\textdagger\newline{\tiny 1934--2023} & University of Toronto & Pole assignment; geometric linear control; internal model principle; supervisory control \citep{wonham1967pole,francis1976internal,ramadge1987supervisory} & Th & Supervisory control of manufacturing and protocols & 5 & 9 & 9 & 4 & 8 & 9 & 70.0\\
43 & \textbf{Boris P. Kovatchev}* & University of Virginia & In-silico testing and closed-loop control for type 1 diabetes \citep{kovatchev2009insilico} & Al/Sw & Algorithm behind a widely used commercial closed-loop insulin system; FDA-accepted simulator & 7 & 6 & 8 & 9 & 6 & 6 & 69.5\\
44 & \textbf{Gunter Stein}\textdagger\newline{\tiny 1941--2026} & Honeywell; MIT (adjunct) & LQG/LTR; multivariable robust flight control \citep{doyle1981multivariable,stein2003respect} & Th/Al & Flight-control systems of advanced aircraft & 7 & 6 & 8 & 7 & 6 & 8 & 69.5\\
45 & \textbf{George Zames}\textdagger\newline{\tiny 1934--1997} & McGill University & Small-gain theorem; $\mathcal{H}_\infty$ formulation of robust control \citep{zames1966input,zames1981feedback} & Th & Conceptual foundation of robust control & 6 & 9 & 9 & 3 & 6 & 8 & 69.5\\
46 & \textbf{Jessy W. Grizzle} [X] & University of Michigan & Hybrid zero dynamics; automotive powertrain control \citep{grizzle2001asymptotically,westervelt2003hybrid} & Th/Hw & Bipedal walking; engine and emissions control & 6 & 7 & 9 & 5 & 8 & 8 & 69.5\\
47 & \textbf{Roger W. Brockett}\textdagger [X]\newline{\tiny 1938--2023} & MIT; Harvard & Geometric control; bilinear systems; Brockett's necessary condition \citep{brockett1970finite,brockett1983asymptotic} & Th/Ed & Theory behind nonholonomic vehicles and attitude control & 5 & 9 & 8 & 3 & 10 & 9 & 69.5\\
48 & \textbf{Tamer Ba\c{s}ar} & UIUC & Dynamic games; game-theoretic $\mathcal{H}_\infty$ control \citep{basar1999dynamic,basar1995hinf} & Th/Ed & Game-theoretic design of networks and secure systems & 5 & 8 & 10 & 3 & 9 & 9 & 69.5\\
49 & \textbf{Elmer G. Gilbert}\textdagger\newline{\tiny 1930--2019} & University of Michigan & Kalman--Gilbert decomposition; GJK distance algorithm; maximal output admissible sets \citep{gilbert1963controllability,gilbert1988fast,gilbert1991linear} & Th/Al & Collision checking in robotics and simulation; constraint management & 6 & 7 & 8 & 6 & 7 & 9 & 69.0\\
50 & \textbf{Michael Athans}\textdagger\newline{\tiny 1937--2020} & MIT & LQG and multivariable design; optimal control textbook \citep{athans1966optimal,athans1971role} & Th/Ed & Multivariable flight and engine control & 6 & 7 & 8 & 5 & 9 & 8 & 69.0\\
51 & \textbf{Jan C. Willems}\textdagger\newline{\tiny 1939--2013} & MIT; Univ. of Groningen; KU Leuven & Dissipativity; behavioural approach; fundamental lemma \citep{willems1972dissipative,willems2005note} & Th & Passivity-based design; foundation of data-driven control & 5 & 9 & 9 & 3 & 8 & 9 & 69.0\\
52 & \textbf{David W. Clarke}\ddag & University of Oxford & Generalised predictive control; self-tuning control \citep{clarke1987generalized} & Al & Predictive control in process industries & 7 & 7 & 8 & 5 & 6 & 7 & 68.5\\
53 & \textbf{Roman Hovorka}* & University of Cambridge & Model predictive closed-loop insulin delivery \citep{hovorka2004nonlinear} & Al/Hw & Commercial hybrid closed-loop insulin delivery (CamAPS FX) & 7 & 6 & 8 & 8 & 6 & 6 & 68.5\\
54 & \textbf{John R. Ragazzini}\textdagger\newline{\tiny 1912--1988} & Columbia University; New York University & Sampled-data control; $z$-transform; operational-amplifier analysis \citep{ragazzini1947analysis,ragazzini1952analysis} & Th/Ed & Digital control; mentor of K\'alm\'an, Zadeh, Jury and Franklin & 6 & 8 & 7 & 4 & 10 & 7 & 68.5\\
55 & \textbf{Sigurd Skogestad} & NTNU & Plantwide and multivariable process control; SIMC tuning rules \citep{skogestad2005multivariable,skogestad2003simple} & Th/Ed & PID tuning and control-structure design in process plants & 6 & 6 & 9 & 6 & 9 & 7 & 68.5\\
56 & \textbf{Francis J. Doyle III} & UC Santa Barbara; Harvard & Zone MPC for automated insulin delivery; systems biology \citep{grosman2010zone,doyle2014closed} & Al/Hw & Closed-loop insulin delivery for type 1 diabetes & 6 & 6 & 9 & 6 & 8 & 8 & 68.5\\
57 & \textbf{Michio Sugeno}\ddag & Tokyo Institute of Technology & Takagi--Sugeno fuzzy models and fuzzy control \citep{takagi1985fuzzy} & Th/Al & Fuzzy control in industry and transport & 6 & 8 & 9 & 4 & 6 & 7 & 67.5\\
58 & \textbf{Petros A. Ioannou} & Univ. of Southern California & Robust adaptive control; intelligent cruise control \citep{ioannou1996robust,ioannou1993autonomous} & Th/Al & Adaptive cruise control and traffic management & 6 & 6 & 9 & 6 & 8 & 7 & 67.5\\
59 & \textbf{Hassan K. Khalil} & Michigan State University & Nonlinear systems; high-gain observers; singular perturbations \citep{khalil2002nonlinear,khalil2014high} & Th/Ed & Observer-based nonlinear design; textbook used worldwide & 5 & 7 & 10 & 3 & 10 & 8 & 67.5\\
60 & \textbf{Suguru Arimoto}\ddag [X] & Osaka Univ.; Ritsumeikan Univ. & Iterative learning control; passivity-based robot control \citep{arimoto1984bettering,takegaki1981new} & Th/Al & Learning control of repetitive manufacturing operations & 6 & 8 & 8 & 4 & 7 & 7 & 67.0\\
61 & \textbf{Arjan van der Schaft} & University of Groningen & Port-Hamiltonian systems; $L_2$-gain and passivity \citep{vanderschaft2017l2,vanderschaft2014port} & Th & Energy-based design of robots and power networks & 5 & 8 & 9 & 3 & 8 & 9 & 67.0\\
61 & \textbf{Bruce A. Francis} & University of Toronto & Internal model principle; $\mathcal{H}_\infty$ control theory \citep{francis1976internal,francis1987course} & Th/Ed & Robust tracking and disturbance rejection & 5 & 8 & 9 & 3 & 8 & 9 & 67.0\\
61 & \textbf{Harold J. Kushner}\ddag & Brown University & Stochastic stability; stochastic approximation; numerical stochastic control \citep{kushner1967stochastic,kushner2003stochastic} & Th & Convergence theory behind stochastic learning algorithms & 5 & 8 & 9 & 3 & 8 & 9 & 67.0\\
64 & \textbf{Frank Allg\"ower} & University of Stuttgart & Nonlinear and data-driven MPC; systems biology \citep{chen1998quasi,berberich2021data} & Th/Al & Predictive and data-driven control methods & 5 & 7 & 9 & 5 & 8 & 9 & 67.0\\
64 & \textbf{Naomi Ehrich Leonard} [X] & Princeton University & Collective motion; coordinated control of underwater glider fleets \citep{leonard2001virtual,leonard2007collective} & Th/Hw & Adaptive ocean sampling with autonomous gliders & 5 & 7 & 9 & 5 & 8 & 9 & 67.0\\
66 & \textbf{Claudio Cobelli} & University of Padova & Glucose--insulin modelling; UVA/Padova simulator for in-silico trials \citep{dallaman2007meal,kovatchev2009insilico} & Th/Sw & Simulator accepted in place of animal trials for insulin-control algorithms & 6 & 6 & 9 & 5 & 8 & 7 & 66.5\\
67 & \textbf{Tianyou Chai} & Northeastern University (China) & Integrated automation and optimal operational control of industrial processes \citep{chai2014optimal} & Al/Hw & Automation of mineral-processing and metallurgy plants & 6 & 6 & 7 & 7 & 8 & 8 & 66.5\\
68 & \textbf{Aleksandr A. Andronov}\textdagger\newline{\tiny 1901--1952} & Gorky State University & Theory of oscillations; structural stability; nonlinear control \citep{andronov1937systemes,andronov1966theory} & Th/Ed & Nonlinear dynamics of regulators and oscillators & 5 & 9 & 8 & 3 & 9 & 7 & 66.5\\
69 & \textbf{Jacques-Louis Lions}\textdagger\newline{\tiny 1928--2001} & Coll\`ege de France; Paris & Optimal control of PDEs; Hilbert uniqueness method \citep{lions1971optimal,lions1988exact} & Th & Control of distributed-parameter systems; space-agency leadership & 3 & 9 & 10 & 4 & 9 & 10 & 66.5\\
70 & \textbf{Aaron D. Ames} [X] & Caltech & Control barrier functions; hybrid-zero-dynamics walking \citep{ames2017control,ames2019control} & Th/Hw & Safety filters for autonomous systems; exoskeleton walking & 5 & 8 & 9 & 5 & 7 & 7 & 66.0\\
71 & \textbf{Claire J. Tomlin} & Stanford; UC Berkeley & Hamilton--Jacobi reachability; hybrid systems; air-traffic conflict resolution \citep{tomlin1998conflict,mitchell2005time} & Th/Al & Safety analysis of aircraft and drone operations & 5 & 7 & 9 & 5 & 8 & 8 & 66.0\\
72 & \textbf{Mark W. Spong} [X] & UIUC; UT Dallas & Elastic-joint robots; passivity-based teleoperation \citep{spong1987modeling,anderson1989bilateral} & Th/Ed & Robot-control practice and education & 5 & 7 & 9 & 4 & 9 & 8 & 66.0\\
73 & \textbf{Yakov Z. Tsypkin}\textdagger\newline{\tiny 1919--1997} & Inst. of Control Sciences; MIPT & Sampled-data and relay systems; adaptation and learning \citep{tsypkin1971adaptation} & Th & Foundations of digital and learning control & 5 & 8 & 8 & 4 & 8 & 8 & 65.5\\
74 & \textbf{Iven Mareels}* & University of Melbourne & Adaptive control; automation of irrigation channel networks [Mareels et al., Proc. IEEE 2005] & Al/Hw & Automated irrigation networks saving water in Australia and abroad & 6 & 5 & 8 & 8 & 7 & 7 & 65.0\\
75 & \textbf{George J. Pappas} & University of Pennsylvania & Hybrid-system abstraction; formal methods for control \citep{alur2000discrete} & Th & Verification of autonomous and cyber-physical systems & 5 & 7 & 9 & 4 & 8 & 8 & 65.0\\
76 & \textbf{John N. Tsitsiklis} & MIT & Neuro-dynamic programming; distributed computation \citep{bertsekas1996neuro,tsitsiklis1986distributed} & Th & Foundations of reinforcement learning and distributed algorithms & 5 & 7 & 9 & 3 & 8 & 9 & 65.0\\
76 & \textbf{P. R. Kumar} & UIUC; Texas A\&M & Stochastic adaptive control; wireless networks; cyber-physical systems \citep{kumar1986stochastic,gupta2000capacity} & Th & Theory of networked and cyber-physical systems & 5 & 7 & 9 & 3 & 8 & 9 & 65.0\\
78 & \textbf{Lorenz T. Biegler}* & Carnegie Mellon University & Dynamic optimisation and nonlinear MPC; interior-point solver IPOPT [W\"achter \& Biegler, Math. Prog. 2006] & Al/Sw & Optimisation-based operation of process plants; solver used across control and robotics & 5 & 6 & 9 & 7 & 8 & 7 & 65.0\\
78 & \textbf{Russ Tedrake} [X] & MIT & Underactuated control; sum-of-squares verification; Drake \citep{tedrake2010lqr,tedrake2019drake} & Al/Sw & Open-source model-based design tools for robots & 5 & 6 & 9 & 7 & 8 & 7 & 65.0\\
80 & \textbf{Francesco Borrelli} & UC Berkeley & MPC for vehicles and energy systems \citep{falcone2007predictive,borrelli2017predictive} & Al & Predictive control in automotive and building systems & 6 & 5 & 9 & 7 & 7 & 6 & 64.5\\
81 & \textbf{Raffaello D'Andrea} [X] & Cornell; ETH Z\"urich & Distributed control of spatially interconnected systems; Kiva multi-robot control \citep{dandrea2003distributed,wurman2008coordinating} & Hw/Th & Coordinated warehouse robots in e-commerce logistics & 6 & 5 & 7 & 9 & 6 & 8 & 64.5\\
82 & \textbf{Vladimir A. Yakubovich}\textdagger\newline{\tiny 1926--2012} & Leningrad State University & Kalman--Yakubovich--Popov lemma; absolute stability \citep{yakubovich1962solution} & Th & Frequency-domain robustness certificates & 5 & 8 & 8 & 3 & 7 & 9 & 64.5\\
83 & \textbf{Ebrahim H. Mamdani}\textdagger\newline{\tiny 1942--2010} & Queen Mary, Univ. of London & First fuzzy-logic controller \citep{mamdani1974application,mamdani1975experiment} & Al & Fuzzy control of kilns and consumer devices & 6 & 8 & 8 & 4 & 5 & 6 & 64.0\\
84 & \textbf{Romeo Ortega} & CNRS--Sup\'elec; ITAM & Passivity-based control; IDA-PBC \citep{ortega1998passivity,ortega2002interconnection} & Th & Control of electromechanical and power-electronic systems & 5 & 8 & 9 & 3 & 7 & 7 & 64.0\\
85 & \textbf{Edward J. Davison}\ddag* & University of Toronto & Robust servomechanism problem; decentralised fixed modes [Davison, IEEE TAC 1976] & Th & Robust tracking design in industrial controllers & 5 & 7 & 9 & 3 & 8 & 8 & 64.0\\
86 & \textbf{Pramod P. Khargonekar} & Univ. of Florida; Univ. of Michigan; UC Irvine & State-space $\mathcal{H}_\infty$ control (DGKF) \citep{doyle1989state} & Th & Robust control design tools & 5 & 7 & 9 & 3 & 7 & 9 & 64.0\\
87 & \textbf{Emilio Frazzoli} [X] & MIT; ETH Z\"urich & Maneuver-based motion planning for agile vehicles \citep{frazzoli2002real} & Al & Autonomous mobility services & 5 & 6 & 9 & 7 & 7 & 7 & 64.0\\
88 & \textbf{Karl Henrik Johansson} & KTH Royal Institute of Technology & Networked and event-triggered control; heavy-vehicle platooning \citep{heemels2012introduction,alam2015heavy} & Th/Hw & Fuel-saving truck platooning; wireless control & 5 & 6 & 9 & 6 & 7 & 8 & 64.0\\
89 & \textbf{Ioan D. Landau}\ddag & Grenoble (CNRS/INP) & Model-reference adaptive control; hyperstability design \citep{landau1979adaptive} & Th/Al & Adaptive control in industry & 5 & 7 & 8 & 6 & 7 & 7 & 63.5\\
89 & \textbf{Michael G. Safonov} & Univ. of Southern California & Multivariable stability margins; unfalsified control; robust-control software \citep{safonov1981feedback,safonov1997unfalsified} & Th/Sw & Robust-control computer-aided design & 5 & 7 & 8 & 6 & 7 & 7 & 63.5\\
91 & \textbf{Pablo A. Parrilo} & MIT & Sum-of-squares programming \citep{parrilo2003semidefinite} & Th/Al/Sw & Polynomial certificates for nonlinear systems & 4 & 8 & 9 & 5 & 7 & 8 & 63.5\\
92 & \textbf{Alistair G. J. MacFarlane}\textdagger\newline{\tiny 1931--2021} & Univ. of Cambridge; UMIST & Characteristic-locus multivariable design \citep{macfarlane1977generalized} & Th/Ed & Computer-aided multivariable design & 5 & 7 & 7 & 5 & 8 & 8 & 63.0\\
93 & \textbf{J\"urgen Ackermann}\ddag* & DLR; TU Munich (adjunct prof.) & Pole-placement formula; parameter-space robust control; robust vehicle steering [Ackermann, book 1993] & Th/Al & Robust steering control demonstrated in road vehicles; eponymous textbook formula & 5 & 6 & 9 & 6 & 6 & 8 & 63.0\\
94 & \textbf{Florian D\"orfler} & ETH Z\"urich & Synchronisation of power grids; data-enabled predictive control \citep{dorfler2013synchronization,coulson2019data} & Th/Al & Stability of inverter-based, low-inertia grids & 5 & 6 & 9 & 5 & 7 & 8 & 63.0\\
95 & \textbf{Eliahu I. Jury}\textdagger\newline{\tiny 1923--2020} & UC Berkeley & Sampled-data stability (Jury test); $z$-transform methods \citep{jury1964theory} & Th & Digital control and filter design & 5 & 7 & 8 & 3 & 8 & 8 & 62.5\\
96 & \textbf{W. Harmon Ray}\ddag* & University of Wisconsin--Madison & Process control; control of polymerisation reactors [Ray, book 1981; Ogunnaike \& Ray, book 1994] & Th/Ed & Control of chemical and polymerisation processes & 5 & 6 & 8 & 5 & 8 & 8 & 62.5\\
97 & \textbf{Gene F. Franklin}\textdagger\newline{\tiny 1927--2012} & Stanford University & Digital and feedback control textbooks \citep{franklin1986feedback,franklin1990digital} & Ed/Th & Textbooks that trained generations of control engineers & 5 & 6 & 8 & 4 & 10 & 7 & 62.5\\
98 & \textbf{Arthur J. Krener}\ddag & UC Davis; Naval Postgraduate School & Nonlinear controllability and observers \citep{hermann1977nonlinear,krener1983linearization} & Th & Nonlinear estimation and design & 4 & 8 & 9 & 3 & 7 & 9 & 62.5\\
99 & \textbf{Charles A. Desoer}\textdagger\newline{\tiny 1926--2010} & UC Berkeley & Input--output theory of feedback systems \citep{desoer1975feedback,zadeh1963linear} & Th/Ed & Rigorous feedback-system analysis; Berkeley school & 4 & 7 & 9 & 3 & 9 & 9 & 62.5\\
100 & \textbf{John B. Moore}\ddag & Australian National University & Optimal filtering and control \citep{anderson1979optimal} & Th/Ed & Estimation and control textbooks & 5 & 6 & 9 & 4 & 8 & 7 & 62.0\\
\end{longtable}
\end{landscape}

\begin{landscape}
\setlength{\LTcapwidth}{\linewidth}
\scriptsize
\setlength{\tabcolsep}{2.6pt}
\renewcommand{\arraystretch}{1.04}
\begin{longtable}{@{}r >{\raggedright\arraybackslash}p{3.15cm} >{\raggedright\arraybackslash}p{3.2cm} >{\raggedright\arraybackslash}p{6.6cm} >{\raggedright\arraybackslash}p{1.4cm} >{\raggedright\arraybackslash}p{4.6cm} cccccc r@{}}
\caption{The top 100 scholars in robotics (1926--2026), ranked by the impact-first composite score. The list has 102 entries because 3 candidates with identical criterion vectors share rank 100.}\label{tab:robotics100}\\
\toprule
\# & Scholar & Key institution(s) & Signature contribution(s) & Nature & Impact on humanity & I & F & S & T & E & R & Score\\
\midrule
\endfirsthead
\multicolumn{13}{@{}l}{\textit{Table \thetable\ (continued)}}\\
\toprule
\# & Scholar & Key institution(s) & Signature contribution(s) & Nature & Impact on humanity & I & F & S & T & E & R & Score\\
\midrule
\endhead
\midrule
\multicolumn{13}{r@{}}{\textit{continued on next page}}\\
\endfoot
\bottomrule
\multicolumn{13}{@{}p{23.85cm}@{}}{\textdagger\ deceased (life span under the name); \ddag\ living status inferred from the absence of any death report as of 27~November~2026 (scholar born in or before 1945, or birth year unverified); [X] in both lists; * added in the revised screening. Nature: Th theory, Al algorithm, Hw hardware/systems, Sw software, Ed education. Criteria (0--10): I impact on humanity (weight 35\%), F field creation (20\%), S scholarly influence (15\%), T technology transfer (10\%), E education and community (10\%), R peer recognition (10\%). Score $=3.5I+2F+1.5S+T+E+R$. Equal scores are ordered by I, then F, S, T, E and R; identical criterion vectors share a rank and are listed alphabetically. The evidence behind every score is in Supplementary Material S1.}\\
\endlastfoot

1 & \textbf{Takeo Kanade} & Carnegie Mellon University & Robot vision; Lucas--Kanade registration; direct-drive arms; Navlab \citep{lucas1981iterative,asada1983design,thorpe1988vision} & Th/Hw/Ed & Vision and autonomous-driving technologies; growth of the CMU Robotics Institute & 9 & 8 & 10 & 8 & 10 & 10 & 90.5\\
2 & \textbf{Rodney A. Brooks} & MIT & Subsumption architecture; behaviour-based robotics \citep{brooks1986robust,brooks1991intelligence} & Th/Hw & Domestic floor-cleaning robots in tens of millions of homes \citep{jones2006robots} & 9 & 9 & 9 & 10 & 8 & 9 & 90.0\\
3 & \textbf{Russell H. Taylor} & IBM Research; Johns Hopkins University & Computer-integrated surgery; ROBODOC; cooperative ``steady-hand'' robots \citep{taylor1994image,taylor2003medical} & Hw/Sw/Ed & Image-guided robotic surgery & 9 & 9 & 9 & 9 & 9 & 9 & 90.0\\
4 & \textbf{Sebastian Thrun} & Carnegie Mellon; Stanford & Probabilistic robotics; Monte Carlo localisation; Stanley \citep{thrun2005probabilistic,dellaert1999monte,thrun2006stanley} & Al/Ed & Seeded the autonomous-driving industry; mass online engineering education & 9 & 8 & 10 & 10 & 8 & 9 & 89.5\\
5 & \textbf{Oussama Khatib} & Stanford University & Artificial potential fields; operational-space formulation; OceanOne \citep{khatib1986real,khatib1987unified,khatib2016ocean} & Th/Al/Hw & Whole-body, human-safe manipulation; deep-sea robotic exploration & 8 & 9 & 9 & 8 & 9 & 9 & 85.5\\
6 & \textbf{Marc H. Raibert} & Carnegie Mellon; MIT & Dynamically balanced legged locomotion \citep{raibert1986legged,raibert2008bigdog} & Hw/Al & Commercial legged robots for inspection and logistics & 8 & 9 & 8 & 10 & 7 & 9 & 84.0\\
7 & \textbf{Hugh Durrant-Whyte} & Univ. of Oxford; Univ. of Sydney & SLAM; decentralised data fusion; autonomous container handling \citep{leonard1991mobile,durrantwhyte2006simultaneous,durrantwhyte2007autonomous} & Al/Hw & Automated container terminals and mining fleets & 8 & 9 & 9 & 8 & 8 & 8 & 83.5\\
8 & \textbf{Lydia E. Kavraki} & Rice University & Probabilistic roadmaps; Open Motion Planning Library \citep{kavraki1996probabilistic,sucan2012open} & Al/Sw & Motion-planning software in industry, research and drug design & 7 & 9 & 9 & 8 & 8 & 9 & 81.0\\
9 & \textbf{Daniela Rus} & Dartmouth; MIT & Self-reconfiguring, distributed and soft robots \citep{rus2001crystalline,rus2015design} & Th/Hw/Ed & Robot autonomy and soft robotics; leadership of MIT CSAIL & 7 & 8 & 9 & 8 & 9 & 9 & 80.0\\
10 & \textbf{William L. Whittaker} & Carnegie Mellon University & Field robotics; nuclear-accident and planetary robots; Boss \citep{urmson2008autonomous} & Hw & Robots for nuclear clean-up, mining and space & 8 & 8 & 7 & 9 & 8 & 8 & 79.5\\
11 & \textbf{Pieter Abbeel} & UC Berkeley & Apprenticeship and deep reinforcement learning for robots \citep{abbeel2004apprenticeship,abbeel2010autonomous} & Al/Ed & Robot-learning companies and open education & 7 & 7 & 10 & 9 & 9 & 8 & 79.5\\
12 & \textbf{J. Kenneth Salisbury} & MIT; Stanford & Articulated hands; PHANToM haptic interface; surgical robotics \citep{salisbury1982articulated,massie1994phantom} & Hw & Haptic interfaces and surgical-robot technology & 8 & 7 & 8 & 9 & 8 & 8 & 79.0\\
13 & \textbf{Ruzena Bajcsy}\ddag & Univ. of Pennsylvania; UC Berkeley & Active perception; GRASP Laboratory; elastic image matching \citep{bajcsy1988active,bajcsy1989multiresolution} & Th/Ed & Machine perception; anatomical atlas registration in medicine & 7 & 9 & 9 & 5 & 9 & 9 & 79.0\\
14 & \textbf{Vijay Kumar} & University of Pennsylvania & Agile aerial robots and swarms; cooperative manipulation \citep{mellinger2011minimum,kushleyev2013towards} & Al/Hw/Ed & Autonomous UAVs for inspection and mapping & 7 & 7 & 9 & 9 & 9 & 9 & 79.0\\
15 & \textbf{Dieter Fox} & University of Washington & Probabilistic localisation; dynamic-window navigation; RGB-D mapping \citep{fox1997dynamic,henry2012rgbd} & Al & Navigation and perception for mobile manipulators & 7 & 7 & 10 & 8 & 8 & 9 & 78.5\\
16 & \textbf{Raffaello D'Andrea} [X] & Cornell; ETH Z\"urich & Mobile-robot order fulfilment (Kiva); aerial robotics \citep{wurman2008coordinating} & Hw & Warehouse robot fleets in e-commerce fulfilment & 8 & 7 & 7 & 10 & 6 & 9 & 77.5\\
17 & \textbf{Neville Hogan} & MIT & Impedance control; MIT-Manus rehabilitation robot \citep{hogan1985impedance,krebs1998robot} & Th/Hw & Robot-aided rehabilitation after stroke \citep{lo2010robot} & 7 & 9 & 8 & 7 & 8 & 8 & 77.5\\
18 & \textbf{Paolo Dario} & Scuola Superiore Sant'Anna & Biorobotics; medical and micro-robotics \citep{dario1996robotics} & Hw/Ed & Medical, assistive and endoscopic devices & 7 & 8 & 8 & 7 & 9 & 9 & 77.5\\
19 & \textbf{Wolfram Burgard} & Univ. of Freiburg; UTN & Probabilistic mapping and localisation; museum tour-guide robots \citep{burgard1999experiences,grisetti2007improved} & Al/Ed & Mapping and navigation methods used in service robots & 7 & 7 & 10 & 7 & 8 & 9 & 77.5\\
20 & \textbf{Roland Siegwart} & EPFL; ETH Z\"urich & Autonomous mobile and aerial robots; robot spin-offs \citep{siegwart2011introduction} & Hw/Ed & Field, inspection and aerial robots; many start-ups & 7 & 6 & 9 & 9 & 9 & 9 & 77.0\\
21 & \textbf{Hugh M. Herr} & MIT & Powered ankle--foot prostheses; biomechatronics \citep{au2009powered,herr2012bionic} & Hw & Powered prostheses that restore natural gait & 8 & 7 & 7 & 10 & 6 & 8 & 76.5\\
22 & \textbf{Shigeo Hirose} & Tokyo Institute of Technology & Snake robots; quadruped walkers; demining robots \citep{hirose1993biologically} & Hw & Inspection, rescue and demining robots & 7 & 8 & 8 & 7 & 8 & 9 & 76.5\\
23 & \textbf{Sergey Levine} & UC Berkeley & End-to-end visuomotor learning; large-scale robot learning \citep{levine2016end,levine2018learning} & Al & Generalist robot policies & 7 & 7 & 10 & 7 & 8 & 8 & 76.5\\
24 & \textbf{Tom\'as Lozano-P\'erez} & MIT & Configuration space; fine-motion planning; task-and-motion planning \citep{lozanoperez1983spatial,lozanoperez1984automatic} & Th/Al & Conceptual foundation of all modern motion planners & 6 & 10 & 9 & 4 & 9 & 9 & 76.5\\
25 & \textbf{Miomir Vukobratovi\'c}\textdagger\newline{\tiny 1931--2012} & University of Belgrade & Zero-moment point; biped and exoskeleton dynamics \citep{vukobratovic1972stability,vukobratovic1974development,vukobratovic2004zero} & Th/Hw & Basis of humanoid walking; early powered exoskeletons & 7 & 10 & 9 & 4 & 7 & 7 & 76.0\\
26 & \textbf{Ken Goldberg} & UC Berkeley & Sensorless manipulation; cloud robotics; Dex-Net grasping \citep{goldberg1993orienting,kehoe2015survey,mahler2017dexnet} & Al/Sw & Robot picking in logistics & 7 & 7 & 9 & 8 & 8 & 8 & 76.0\\
27 & \textbf{Yoshiyuki Sankai} & University of Tsukuba & HAL wearable exoskeleton driven by bioelectrical signals \citep{suzuki2007intention} & Hw & Clinical exoskeleton therapy & 8 & 7 & 7 & 10 & 6 & 7 & 75.5\\
28 & \textbf{Gerd Hirzinger}\ddag & DLR (hon. prof. TU Munich) & Lightweight torque-controlled arms; ROTEX space telerobotics \citep{hirzinger1993sensor,albuschaffer2007dlr} & Hw & Collaborative lightweight arms; space robotics & 7 & 7 & 8 & 9 & 7 & 9 & 75.5\\
29 & \textbf{Jean-Claude Latombe} & Stanford University & Robot motion planning; probabilistic roadmaps \citep{latombe1991robot,kavraki1996probabilistic} & Al/Ed & Planning in robotics, animation and biology & 6 & 9 & 9 & 5 & 9 & 9 & 75.5\\
30 & \textbf{Richard P. Paul}\textdagger\newline{\tiny 1936/37--2024} & Stanford; Purdue; Univ. of Pennsylvania & Homogeneous-transform kinematics; robot programming and control \citep{paul1972modelling,paul1981robot} & Th/Sw & Kinematic and programming conventions of industrial robots & 7 & 8 & 8 & 6 & 8 & 8 & 74.5\\
31 & \textbf{Zexiang Li} & Hong Kong Univ. of Science and Technology & Nonholonomic manipulation planning; robotics textbook; robotics entrepreneurship \citep{murray1994mathematical,li1990motion} & Th/Ed & Consumer and agricultural drones through the DJI ecosystem & 7 & 5 & 8 & 10 & 9 & 9 & 74.5\\
32 & \textbf{Matthew T. Mason} & Carnegie Mellon University & Mechanics of manipulation; pushing; compliance \citep{mason1981compliance,mason1986mechanics,mason2001mechanics} & Th & Manipulation principles used in automation & 6 & 9 & 9 & 4 & 9 & 9 & 74.5\\
33 & \textbf{Sami Haddadin} & TU Munich; MBZUAI & Safe physical human--robot interaction; collision analysis \citep{haddadin2009requirements,haddadin2017robot} & Hw/Th & Collaborative-robot safety and tactile arms & 7 & 6 & 9 & 9 & 7 & 8 & 74.0\\
34 & \textbf{Ernst D. Dickmanns}\ddag & Bundeswehr University Munich & Dynamic-vision autonomous driving \citep{dickmanns1987autonomous,dickmanns2007dynamic} & Al/Hw & Origins of vision-based autonomous road vehicles & 7 & 9 & 8 & 5 & 7 & 7 & 73.5\\
35 & \textbf{Steven M. LaValle} & UIUC; Univ. of Oulu & Rapidly-exploring random trees; planning algorithms \citep{lavalle2001randomized,lavalle2006planning} & Al/Ed & Sampling-based planning across robotics and vehicles & 6 & 8 & 9 & 7 & 9 & 7 & 73.5\\
36 & \textbf{Russ Tedrake} [X] & MIT & Underactuated robotics; Drake; robot learning \citep{collins2005efficient,tedrake2010lqr,tedrake2019drake} & Al/Sw & Open-source simulation and control software for robots & 6 & 7 & 9 & 8 & 9 & 8 & 73.5\\
37 & \textbf{Hiroshi Makino}\ddag & University of Yamanashi & SCARA arm \citep{makino2014development} & Hw & SCARA arms in electronics assembly worldwide & 8 & 8 & 6 & 9 & 5 & 6 & 73.0\\
38 & \textbf{Ichiro Kato}\textdagger\newline{\tiny 1925--1994} & Waseda University & WABOT-1 full-scale humanoid; WL biped series \citep{kato1974information,takanishi1985realization} & Hw & Founded humanoid robotics & 7 & 9 & 7 & 4 & 9 & 7 & 73.0\\
39 & \textbf{Andrew J. Davison} & Imperial College London & Real-time monocular SLAM \citep{davison2007monoslam} & Al & Visual SLAM in consumer robots and augmented reality & 7 & 7 & 9 & 6 & 7 & 8 & 73.0\\
40 & \textbf{Stephen C. Jacobsen}\textdagger\newline{\tiny 1940--2016} & University of Utah & Utah Arm prosthesis; Utah/MIT dextrous hand; exoskeletons \citep{jacobsen1982development,jacobsen1986design} & Hw & Myoelectric prostheses and teleoperated robots & 7 & 7 & 7 & 9 & 7 & 8 & 73.0\\
41 & \textbf{Marco Hutter} & ETH Z\"urich & ANYmal quadruped; learned legged locomotion \citep{hutter2016anymal,hwangbo2019learning,miki2022learning} & Hw/Al & Legged inspection robots in industry & 7 & 6 & 9 & 9 & 7 & 7 & 73.0\\
42 & \textbf{Homayoon Kazerooni} & UC Berkeley & Lower-extremity exoskeletons \citep{kazerooni2005control,zoss2006biomechanical} & Hw & Medical and industrial exoskeletons & 7 & 7 & 8 & 9 & 6 & 7 & 72.5\\
43 & \textbf{Cynthia Breazeal} & MIT & Social robots \citep{breazeal2002designing} & Hw/Ed & Social robotics and AI education & 6 & 8 & 9 & 6 & 8 & 8 & 72.5\\
44 & \textbf{Bernard Roth}\ddag & Stanford University & Manipulator kinematics and inverse kinematics \citep{pieper1969kinematics} & Th/Ed & Kinematics underlying industrial arms & 6 & 8 & 8 & 5 & 9 & 9 & 72.0\\
45 & \textbf{Robert J. Wood} & Harvard University & Insect-scale flying robots; pop-up fabrication; soft robots \citep{wood2008first,ma2013controlled} & Hw & Micro-fabrication and soft robotics & 6 & 8 & 9 & 6 & 7 & 8 & 71.5\\
46 & \textbf{Maja J. Matari\'c} & Univ. of Southern California & Behaviour-based multi-robot systems; socially assistive robotics \citep{mataric1992integration,feilseifer2005defining} & Th/Hw & Robots supporting therapy for autism, stroke and dementia & 6 & 8 & 9 & 4 & 9 & 8 & 71.5\\
46 & \textbf{Thomas B. Sheridan}\ddag & MIT & Supervisory control of telerobots; human--automation interaction \citep{sheridan1978human,sheridan1992telerobotics} & Th & Levels-of-automation thinking behind teleoperation and autonomy & 6 & 8 & 9 & 4 & 9 & 8 & 71.5\\
48 & \textbf{Dario Floreano} & EPFL & Evolutionary and aerial robotics \citep{nolfi2000evolutionary,floreano2015science} & Hw/Al & Fixed-wing mapping drones & 6 & 7 & 9 & 8 & 8 & 7 & 71.5\\
49 & \textbf{Antonio Bicchi} & Univ. of Pisa; IIT & Variable-impedance actuation; soft synergy hands \citep{bicchi2004fast,catalano2014adaptive} & Th/Hw & Soft prosthetic and robotic hands & 6 & 7 & 9 & 6 & 8 & 9 & 71.5\\
50 & \textbf{Reymond Clavel} & EPFL & Delta parallel robot \citep{clavel1988delta} & Hw & High-speed pick-and-place in food, pharma and electronics & 7 & 8 & 7 & 9 & 5 & 6 & 71.0\\
51 & \textbf{Chelsea Finn} & Stanford University & Meta-learning; generalist robot policies \citep{finn2017model,brohan2023rt2} & Al & Foundation models for robot manipulation & 6 & 7 & 10 & 6 & 7 & 8 & 71.0\\
52 & \textbf{Jean-Jacques E. Slotine} [X] & MIT & Adaptive and sliding control of manipulators; wave-variable teleoperation \citep{slotine1987adaptive,niemeyer1991stable} & Th & Robust robot and teleoperator control & 6 & 8 & 9 & 5 & 8 & 7 & 70.5\\
53 & \textbf{Emilio Frazzoli} [X] & MIT; ETH Z\"urich & Asymptotically optimal sampling-based planning (RRT*); mobility on demand \citep{karaman2011sampling,pavone2012robotic} & Al & Autonomous mobility-on-demand services & 6 & 7 & 9 & 8 & 7 & 7 & 70.5\\
54 & \textbf{Allison M. Okamura} & Johns Hopkins; Stanford & Haptics; needle steering; growing vine robots \citep{webster2006nonholonomic,okamura2009haptic,hawkes2017soft} & Hw/Th & Haptic feedback in surgery; soft growing robots & 6 & 7 & 9 & 5 & 8 & 9 & 70.5\\
54 & \textbf{John J. Leonard} & MIT & Simultaneous localisation and mapping; marine robotics \citep{leonard1991mobile,cadena2016past} & Al/Hw & Autonomous navigation of marine and road vehicles & 6 & 7 & 9 & 5 & 8 & 9 & 70.5\\
56 & \textbf{Guang-Zhong Yang} & Imperial College London; Shanghai Jiao Tong Univ. & Medical robotics; founding editor of \emph{Science Robotics} \citep{yang2017medical,yang2018grand} & Hw/Ed & Surgical robotics and community building & 6 & 6 & 9 & 6 & 9 & 9 & 70.5\\
57 & \textbf{Robert Riener} & ETH Z\"urich & Patient-cooperative gait rehabilitation; Cybathlon \citep{riener2005patient,riener2016cybathlon} & Hw/Ed & Robot-assisted gait therapy & 7 & 6 & 8 & 7 & 7 & 7 & 69.5\\
58 & \textbf{Frank Dellaert} & Georgia Tech & Factor graphs and smoothing; Monte Carlo localisation \citep{dellaert1999monte,dellaert2006square,dellaert2017factor} & Al/Sw & Factor-graph estimation in robots and AR devices & 6 & 7 & 9 & 7 & 7 & 7 & 69.5\\
59 & \textbf{Bradley J. Nelson} & ETH Z\"urich & Micro- and nano-robotics for medicine \citep{nelson2010microrobots} & Hw & Magnetically guided medical microrobots & 6 & 7 & 9 & 6 & 7 & 8 & 69.5\\
60 & \textbf{Hirochika Inoue}\ddag & University of Tokyo & Hand--eye systems; remote-brained and humanoid robots \citep{inaba2000platform} & Hw/Ed & Japan's humanoid-robot programme & 6 & 7 & 7 & 6 & 9 & 9 & 69.5\\
61 & \textbf{Hans P. Moravec} & Stanford; Carnegie Mellon & Stanford Cart; occupancy grids \citep{moravec1983stanford,moravec1985high} & Th/Hw & Occupancy-grid mapping; vision-guided industrial vehicles & 6 & 8 & 8 & 7 & 6 & 7 & 69.0\\
62 & \textbf{Toshio Fukuda} & Nagoya University & Cellular (reconfigurable) and micro/nano robotics \citep{fukuda1988dynamically} & Th/Hw & Micro- and nano-manipulation & 5 & 8 & 9 & 5 & 8 & 9 & 69.0\\
63 & \textbf{Robert W. Mann}\textdagger\newline{\tiny 1924--2006} & MIT & Boston Arm myoelectric prosthesis; sensory aids for the blind \citep{mit2006mann} & Hw/Ed & Early myoelectric arm; Braille translation technology & 7 & 7 & 6 & 6 & 8 & 7 & 68.5\\
64 & \textbf{Jacques Denavit \& Richard S. Hartenberg (joint)}\textdagger\newline{\tiny Denavit 1930--2013; Hartenberg deceased} & Northwestern University & Denavit--Hartenberg kinematic notation \citep{denavit1955kinematic,hartenberg1964kinematic} & Th/Ed & Kinematic convention used in robot controllers and courses & 6 & 8 & 9 & 5 & 8 & 5 & 68.5\\
65 & \textbf{Cl\'ement Gosselin} & Universit\'e Laval & Parallel mechanisms; underactuated hands \citep{gosselin1990singularity,birglen2008underactuated} & Th/Hw & Adaptive grippers used in industry & 6 & 6 & 9 & 8 & 7 & 7 & 68.5\\
66 & \textbf{Martial Hebert} & Carnegie Mellon University & 3-D perception for autonomous vehicles \citep{thorpe1988vision} & Al/Ed & Perception for field and road robots & 6 & 6 & 9 & 5 & 9 & 8 & 68.5\\
67 & \textbf{Leslie Pack Kaelbling} & MIT & Planning under partial observability; integrated task and motion planning \citep{kaelbling1998planning,kaelbling2011hierarchical} & Th/Al & Planning methods for autonomous robots & 5 & 8 & 10 & 4 & 8 & 8 & 68.5\\
68 & \textbf{Suguru Arimoto}\ddag [X] & Osaka Univ.; Ritsumeikan Univ. & PD-plus-gravity stability; iterative learning of robot motion \citep{takegaki1981new,arimoto1984bettering} & Th & Simple, provably stable control of industrial robots & 6 & 8 & 8 & 4 & 7 & 8 & 68.0\\
69 & \textbf{Daniel E. Whitney}\ddag & MIT; Draper Laboratory & Resolved-motion-rate control; remote-centre compliance \citep{whitney1969resolved,whitney1982quasi} & Th/Hw & Compliant assembly devices in manufacturing & 6 & 7 & 8 & 6 & 7 & 8 & 68.0\\
70 & \textbf{Daniel E. Koditschek} & Univ. of Michigan; Univ. of Pennsylvania & Navigation functions; RHex; templates and anchors \citep{rimon1992exact,saranli2001rhex,full1999templates} & Th/Hw & Design principles of legged robots & 5 & 8 & 9 & 5 & 8 & 8 & 68.0\\
70 & \textbf{Jessy W. Grizzle} [X] & University of Michigan & Hybrid-zero-dynamics walking (MABEL, Cassie) \citep{westervelt2003hybrid,westervelt2007feedback} & Th/Hw & Agile bipedal robots & 5 & 8 & 9 & 5 & 8 & 8 & 68.0\\
72 & \textbf{Hermano Igo Krebs} & MIT & Rehabilitation robots (MIT-Manus/InMotion) \citep{krebs1998robot,lo2010robot} & Hw & Robot-assisted stroke therapy & 7 & 6 & 8 & 7 & 6 & 6 & 67.5\\
73 & \textbf{Bruno Siciliano} & University of Naples Federico II & Robot modelling and control; \emph{Springer Handbook of Robotics} \citep{siciliano2009robotics,siciliano2016springer} & Th/Ed & Reference works that structure robotics education & 5 & 6 & 10 & 4 & 10 & 9 & 67.5\\
74 & \textbf{Gill A. Pratt} & MIT; Olin College & Series elastic actuators \citep{pratt1995series} & Hw & Compliant actuators in collaborative and legged robots & 6 & 7 & 8 & 6 & 7 & 7 & 67.0\\
75 & \textbf{Charles E. Thorpe} & Carnegie Mellon University & Navlab autonomous vehicles \citep{thorpe1988vision} & Hw/Al & Early autonomous road vehicles & 6 & 7 & 8 & 5 & 8 & 7 & 67.0\\
76 & \textbf{Howie Choset} & Carnegie Mellon University & Snake robots; coverage planning \citep{choset2001coverage,choset2005principles} & Hw/Al & Surgical and inspection snake robots & 6 & 6 & 8 & 7 & 8 & 7 & 67.0\\
77 & \textbf{Manuela M. Veloso} & Carnegie Mellon University & Multi-robot soccer (RoboCup); symbiotic service robots \citep{kitano1997robocup,veloso2015cobots} & Al & Autonomous service robots; robot competitions & 5 & 7 & 9 & 5 & 8 & 9 & 67.0\\
77 & \textbf{Mark R. Cutkosky} & Stanford University & Grasp taxonomy; bio-inspired adhesion and climbing \citep{cutkosky1989grasp,kim2008smooth} & Hw & Grasp design; gecko-inspired adhesives & 5 & 7 & 9 & 5 & 8 & 9 & 67.0\\
79 & \textbf{Shuuji Kajita}* & AIST; Chubu University (professor) & Linear inverted pendulum; preview control of the zero-moment point \citep{kajita2003biped} & Al/Th & Walking-pattern generation used across humanoid robots & 6 & 7 & 9 & 5 & 6 & 7 & 66.5\\
80 & \textbf{Davide Scaramuzza} & University of Zurich & Event-based vision; agile vision-based drones \citep{gallego2022event,kaufmann2023champion} & Al & Vision-based drone navigation & 6 & 6 & 9 & 6 & 7 & 7 & 66.5\\
81 & \textbf{Stefan Schaal} & Univ. of Southern California; MPI & Movement primitives; learning control \citep{schaal1999imitation,ijspeert2013dynamical} & Al & Robot learning and industrial robot software & 5 & 7 & 10 & 6 & 7 & 7 & 66.5\\
82 & \textbf{Cecilia Laschi} & Scuola Superiore Sant'Anna; National Univ. of Singapore & Octopus-inspired soft robotics \citep{laschi2012soft,laschi2016soft} & Hw & Founding soft robotics as a field & 5 & 8 & 9 & 4 & 8 & 7 & 66.0\\
83 & \textbf{Marco Dorigo} & Universit\'e Libre de Bruxelles & Swarm intelligence and swarm robotics \citep{dorigo1996ant,brambilla2013swarm} & Al & Swarm-engineering methods & 5 & 8 & 9 & 4 & 7 & 8 & 66.0\\
84 & \textbf{Aude Billard} & EPFL & Learning from demonstration; dynamical-system motion control \citep{billard2008robot,khansarizadeh2011learning} & Al & Teaching robots by demonstration & 5 & 7 & 9 & 5 & 8 & 8 & 66.0\\
84 & \textbf{Naomi Ehrich Leonard} [X] & Princeton University & Coordinated autonomous underwater gliders \citep{leonard2007collective} & Th/Hw & Autonomous ocean sampling & 5 & 7 & 9 & 5 & 8 & 8 & 66.0\\
86 & \textbf{Mark W. Spong} [X] & UIUC; UT Dallas & Elastic-joint and passivity-based robot control; robot-control textbook \citep{spong1987modeling,spong2006robot} & Th/Ed & Robot-control practice and education & 5 & 7 & 9 & 4 & 9 & 8 & 66.0\\
86 & \textbf{Seth Hutchinson} & UIUC; Georgia Tech & Visual servo control \citep{hutchinson1996tutorial,chaumette2006visual} & Th/Ed & Vision-guided robot control & 5 & 7 & 9 & 4 & 9 & 8 & 66.0\\
88 & \textbf{Yoshihiko Nakamura} & University of Tokyo & Redundancy resolution; humanoid dynamics \citep{nakamura1987task,nakamura1991advanced} & Th & Control of redundant manipulators and humanoids & 5 & 7 & 9 & 4 & 8 & 9 & 66.0\\
89 & \textbf{Juan D. Tard\'os} & Universidad de Zaragoza & ORB-SLAM \citep{murartal2015orb} & Al/Sw & Open-source visual SLAM & 6 & 6 & 9 & 6 & 7 & 6 & 65.5\\
90 & \textbf{Ferdinand Freudenstein}\textdagger\newline{\tiny 1926--2006} & Columbia University & Kinematic synthesis of mechanisms \citep{freudenstein1955approximate} & Th/Ed & Mechanism design underlying robot kinematics & 5 & 8 & 8 & 3 & 9 & 8 & 65.5\\
91 & \textbf{Jean-Paul Laumond}\textdagger*\newline{\tiny 1953--2021} & LAAS-CNRS; Coll\`ege de France (annual chair 2011--2012) & Nonholonomic motion planning; probabilistic planning software [Laumond, IJCAI 1987; book 1998] & Al/Sw & Path-planning software in industrial CAD (Kineo CAM, acquired by Siemens) & 5 & 7 & 8 & 8 & 6 & 8 & 65.5\\
92 & \textbf{H. Harry Asada} & MIT & Direct-drive arms; robot analysis and design \citep{asada1983design,asada1986robot} & Hw/Th & Direct-drive and assistive robot designs & 5 & 7 & 8 & 6 & 8 & 8 & 65.5\\
93 & \textbf{George A. Bekey}\textdagger\newline{\tiny 1928--2024} & Univ. of Southern California & Biologically inspired autonomous robots; robot ethics \citep{bekey2005autonomous,lin2012robot} & Th/Ed & Autonomous-robotics education and ethics & 5 & 7 & 8 & 4 & 9 & 9 & 65.5\\
94 & \textbf{Susumu Tachi} & University of Tokyo & Telexistence; guide-dog robot \citep{tachi2010telexistence} & Th/Hw & Telepresence and assistive robotics & 5 & 8 & 7 & 6 & 7 & 8 & 65.0\\
95 & \textbf{Peter Corke} & Queensland Univ. of Technology & Robotics Toolbox; robotics and vision textbook \citep{corke1996robotics,corke2017robotics} & Sw/Ed & Global robotics education & 5 & 5 & 9 & 7 & 10 & 7 & 65.0\\
96 & \textbf{Brian L. Davies}\ddag & Imperial College London & Early active surgical robots (Probot, Acrobot) \citep{davies1991development} & Hw & Robot-assisted prostate and orthopaedic surgery & 6 & 7 & 7 & 7 & 6 & 6 & 64.5\\
97 & \textbf{Masahiro Mori}\textdagger\newline{\tiny 1927--2025} & Tokyo Institute of Technology & Uncanny valley; robot contests \citep{mori1970bukimi,mori2012uncanny} & Th/Ed & Design of humanoid and social robots & 5 & 8 & 8 & 4 & 8 & 7 & 64.5\\
98 & \textbf{Roger W. Brockett}\textdagger [X]\newline{\tiny 1938--2023} & Harvard University & Nonholonomic stabilisation; robot-control theory \citep{brockett1983asymptotic} & Th & Theory of wheeled-robot and manipulation control & 5 & 8 & 8 & 3 & 8 & 8 & 64.5\\
99 & \textbf{Joel W. Burdick} & Caltech & Hyper-redundant robots; spinal-cord neuroprostheses \citep{chirikjian1994modal,harkema2011effect} & Th/Hw & Epidural stimulation research for paralysis & 6 & 6 & 8 & 4 & 8 & 7 & 64.0\\
100 & \textbf{Aaron D. Ames} [X] & Caltech & Bipedal and exoskeleton walking; safety-critical robot control \citep{ames2017control} & Th/Hw & Exoskeleton gait and safety filters & 5 & 7 & 9 & 5 & 7 & 7 & 64.0\\
100 & \textbf{Hiroshi Ishiguro} & Osaka University & Android robots; human--robot interaction \citep{kanda2004interactive} & Hw & Social and teleoperated androids & 5 & 7 & 9 & 5 & 7 & 7 & 64.0\\
100 & \textbf{Metin Sitti} & Carnegie Mellon; MPI-IS; Ko\c{c} University & Small-scale medical robots \citep{sitti2015biomedical} & Hw & Untethered milli/micro-robots & 5 & 7 & 9 & 5 & 7 & 7 & 64.0\\
\end{longtable}
\end{landscape}

\stopwide

\section{Honourable mentions}\label{app:hm}
\setcounter{table}{0}

Table~\ref{tab:hm} lists the twenty candidates that follow each list under the baseline weights, with their inclusion frequencies under the weight and score perturbation models of Section~\ref{sec:sensitivity}.

\begin{table*}[!t]
\centering
\caption{Honourable mentions: the twenty candidates that follow each list under the baseline weights. $W_{100}$ and $S_{100}$ are inclusion frequencies (share of simulations in which the candidate reaches rank 100 or better) under random impact-first re-weighting and under $\pm1$ perturbation of every criterion score, respectively (Section~\ref{sec:sensitivity}). They describe sensitivity to these modelling choices, not a probability of belonging to a historical top 100.}
\label{tab:hm}
\scriptsize
\setlength{\tabcolsep}{2.6pt}
\begin{tabular}{@{}r >{\raggedright\arraybackslash}p{2.6cm} >{\raggedright\arraybackslash}p{3.2cm} r r r @{\hspace{8pt}} r >{\raggedright\arraybackslash}p{2.6cm} >{\raggedright\arraybackslash}p{3.2cm} r r r@{}}
\toprule
\multicolumn{6}{c}{Control systems} & \multicolumn{6}{c}{Robotics}\\
\cmidrule(r){1-6}\cmidrule(l){7-12}
\# & Scholar & Principal contribution & Score & $W_{100}$ & $S_{100}$ & \# & Scholar & Principal contribution & Score & $W_{100}$ & $S_{100}$\\
\midrule
101 & Beh\c{c}et A\c{c}{\i}kme\c{s}e & Lossless convexification for powered-descent guidance & 61.5 & 0.42 & 0.47 & 103 & Christopher G. Atkeson & Locally weighted learning for robot control & 64.0 & 0.42 & 0.52\\
102 & Anuradha M. Annaswamy & Stable adaptive systems & 61.5 & 0.40 & 0.48 & 103 & Ronald C. Arkin & Behaviour-based architectures & 64.0 & 0.42 & 0.53\\
103 & Wendell H. Fleming\textdagger & Stochastic optimal control & 61.5 & 0.32 & 0.47 & 105 & John F. Canny & Complexity of robot motion planning & 64.0 & 0.40 & 0.52\\
104 & A. Stephen Morse & Adaptive and supervisory control & 61.5 & 0.30 & 0.47 & 106 & Nicholas Roy & Planning under uncertainty & 64.0 & 0.44 & 0.51\\
104 & Mathukumalli Vidyasagar & Factorisation approach to feedback & 61.5 & 0.30 & 0.47 & 106 & Robert D. Howe & Tactile sensing & 64.0 & 0.44 & 0.52\\
106 & Mustafa Khammash & Cybergenetics & 61.0 & 0.15 & 0.42 & 108 & Jonathan W. Hurst & Spring-mass bipedal robots (ATRIAS, Cassie) & 63.5 & 0.44 & 0.47\\
107 & Nikolai N. Krasovskii\textdagger & Lyapunov--Krasovskii functionals & 61.0 & 0.22 & 0.42 & 109 & Robert J. Full & Biomechanics-inspired robot design & 63.5 & 0.21 & 0.46\\
108 & Sanjoy K. Mitter\textdagger & Nonlinear filtering & 61.0 & 0.20 & 0.42 & 110 & Sangbae Kim & Proprioceptive actuation & 63.0 & 0.30 & 0.41\\
109 & Vasile M. Popov & Popov criterion & 60.5 & 0.23 & 0.35 & 111 & Auke J. Ijspeert & Central pattern generators & 63.0 & 0.02 & 0.41\\
110 & Bart De Moor* & Subspace system identification (N4SID) & 60.5 & 0.13 & 0.36 & 111 & Kerstin Dautenhahn & Social robotics for autism therapy & 63.0 & 0.02 & 0.41\\
111 & Christos G. Cassandras* & Discrete-event systems & 60.5 & 0.02 & 0.37 & 113 & Marco Pavone & Risk-aware autonomy & 63.0 & 0.11 & 0.40\\
112 & Andrew R. Teel & Hybrid dynamical systems & 60.5 & 0.04 & 0.36 & 114 & Richard M. Murray & Mathematical robotic manipulation & 63.0 & 0.17 & 0.41\\
112 & David J. Hill* & Dissipative systems (Hill--Moylan) & 60.5 & 0.04 & 0.36 & 115 & Alessandro De Luca & Flexible robots & 63.0 & 0.14 & 0.41\\
114 & Magnus Egerstedt & Graph-theoretic multi-agent control & 60.5 & 0.08 & 0.36 & 116 & Kristin Y. Pettersen* & Snake robots on land and under water & 62.5 & 0.12 & 0.35\\
115 & Dragoslav D. \v{S}iljak* & Decentralised control of large-scale systems & 60.0 & 0.03 & 0.31 & 117 & Masayuki Inaba* & Remote-brained and humanoid robots (JSK laboratory) & 62.5 & 0.08 & 0.35\\
116 & Frank L. Lewis & Neural-network and reinforcement-learning-based feedback control & 60.0 & 0.08 & 0.27 & 118 & Radhika Nagpal & Programmable self-assembly in robot swarms & 62.5 & 0.01 & 0.35\\
117 & Lino Guzzella* & Control of engines and hybrid powertrains & 59.5 & 0.14 & 0.26 & 119 & Siddhartha S. Srinivasa & Manipulation planning & 62.0 & 0.00 & 0.30\\
118 & A. Galip Ulsoy* & Control of manufacturing and automotive systems & 59.5 & 0.05 & 0.26 & 120 & Blake Hannaford & Teleoperation and haptics & 61.5 & 0.01 & 0.25\\
119 & Thomas F. Edgar* & Process control and optimisation & 59.5 & 0.18 & 0.25 & 121 & Atsuo Takanishi & WABIAN humanoids & 61.5 & 0.00 & 0.24\\
120 & Laurent El Ghaoui* & Linear matrix inequalities in control & 59.5 & 0.02 & 0.26 & 122 & J. Y. S. Luh* & Recursive Newton--Euler manipulator dynamics & 61.0 & 0.01 & 0.21\\
\bottomrule
\multicolumn{12}{@{}l}{\footnotesize * added in the revised screening; \textdagger\ deceased.}\\
\end{tabular}
\end{table*}

\section{Identification and screening protocol}\label{app:protocol}
\setcounter{table}{0}

\paragraph{Rosters} All recipients of six rosters were screened: the IEEE Control Systems Award (1982-2027), the AACC Richard E. Bellman Control Heritage Award (1979-2026), the IFAC Giorgio Quazza Medal (1981-2023) and the IEEE CSS Hendrik W. Bode Lecture Prize (1989-2025) for control, and the IEEE Robotics and Automation Award (2004-2027) and the IEEE RAS Pioneer Award (1999-2026) for robotics. Two further rosters, the Engelberger Robotics Award (technology and education categories) and the IEEE Medal of Honor, were used only as recognition evidence. Each roster is transcribed with its source URL in \texttt{rosters.py}; recipients are matched to candidates by normalised name, and the build script stops if any recipient of an identification roster is neither scored nor in the screening log.

\paragraph{Targeted search} Further names were collected from the histories, handbooks, roadmaps, obituaries and citation databases listed in Section~\ref{sec:screening} and from the author's seed lists, with deliberate searches for the Soviet and Russian, Japanese, Chinese and process-control schools and for women scholars.

\paragraph{Screening} Each identified name was checked against the eligibility rule, the time window and the field scope of Section~\ref{sec:scope}, in that order, and either scored or logged with code P, A or B (Table~\ref{tab:screening}). For appointment checks (code A, and every scored candidate whose qualifying post is not an ordinary faculty appointment) the source consulted is recorded. Names whose principal contribution lies in another field but who also contributed to control or robotics were scored if a field-relevant contribution could be documented, and logged as B otherwise.

\paragraph{Scoring} Each scored candidate received six integer scores against the anchors of Table~\ref{tab:rubric}, with S assessed within decade cohorts, and an evidence entry for each criterion (S1). Scores were proposed with AI assistance and reviewed by the author (Section~\ref{sec:aiuse}); the near-cut-off candidates are subject to independent reassessment (Section~\ref{sec:reassess}).

\begin{table*}[!t]
\centering
\caption{Screening log: every identified name that was not scored, with its rule-based exclusion code and reason. Sources for appointment checks are listed in \texttt{screening\_log.csv}. No name was excluded on a judgement about its likely rank.}
\label{tab:screening}
\scriptsize
\begin{tabular}{@{}l >{\raggedright\arraybackslash}p{3.6cm} >{\raggedright\arraybackslash}p{12.6cm}@{}}
\toprule
Field & Code & Names (reason)\\
\midrule
Control & P: precursor (principal contribution before 1926) & James Clerk Maxwell (On governors (1868)); Edward J. Routh (Stability criterion (1877)); Adolf Hurwitz (Stability criterion (1895)); Ivan A. Vyshnegradsky (Governor theory (1876--1877)); Aleksandr M. Lyapunov (Stability of motion (1892)); Nicolas Minorsky (Three-term ship steering (1922))\\
 & A: no qualifying academic appointment found & Harry Nyquist (Career at AT\&T and Bell Laboratories); Harold S. Black (Career at Bell Laboratories); John C. Lozier (Career at Bell Laboratories); John G. Ziegler (Career at Taylor Instruments); Nathaniel B. Nichols (Taylor Instruments, MIT Radiation Laboratory staff, Aerospace Corporation; no faculty title found); Walter R. Evans (Career at Autonetics and Ford Aeronutronic; no faculty title found); Harold Chestnut (Career at General Electric); Stanley F. Schmidt (NASA Ames Research Center); Jacques Richalet (Industrial career (Adersa)); Paul J. Werbos (National Science Foundation programme director); Michel Fliess (CNRS research director; no faculty title found); Aleksandr A. Feldbaum (Institute of Automation and Telemechanics; no documented teaching post found)\\
 & B: principal contribution outside both fields & Bradford W. Parkinson (Principal contribution is satellite-navigation system architecture (GPS)); Richard S. Sutton (Artificial intelligence (reinforcement learning)); Andrew G. Barto (Artificial intelligence (reinforcement learning)); Ronald A. Howard (Decision analysis and operations research); Boris T. Polyak (Mathematical optimisation); Arkadi Nemirovski (Mathematical optimisation); Yurii Nesterov (Mathematical optimisation); Frank P. Kelly (Communication networks (congestion control)); Steven H. Low (Communication networks (congestion control)); Van Jacobson (Internet protocols (also no academic appointment))\\
\midrule
Robotics & A: no qualifying academic appointment found & George C. Devol (Inventor and entrepreneur (Unimation)); Joseph F. Engelberger (Industrial career (Unimation)); Seiuemon Inaba (Industrial career (FANUC)); Frank Wang (Industrial career (DJI)); Victor Scheinman (Stanford research staff and industry; no faculty title found); John J. Craig (Industry and short-course teaching; no faculty title found); Paul G. Backes (NASA Jet Propulsion Laboratory); Larry H. Matthies (NASA Jet Propulsion Laboratory (adjunct titles not found)); Charles A. Rosen (SRI International); Georges Giralt (LAAS-CNRS research director; no faculty title found)\\
 & B: principal contribution outside both fields & Peter B. Luh (Manufacturing scheduling and power-system optimisation (automation, not robotics)); Nils J. Nilsson (Artificial intelligence (Shakey's planner is discussed in Section 4)); Marvin Minsky (Artificial intelligence); John McCarthy (Artificial intelligence); Raj Reddy (Artificial intelligence (founded the CMU Robotics Institute))\\
\bottomrule
\end{tabular}
\end{table*}

\section{Pioneers not ranked}\label{app:context}
\setcounter{table}{0}

Table~\ref{tab:context} lists influential figures excluded by the time-window, eligibility or field-scope rules of Section~\ref{sec:scope}. Their exclusion reflects the scope of this review, not a judgement of their importance.

\begin{table*}[!t]
\centering
\caption{Influential pioneers excluded by the time-window (P), eligibility (A) or field-scope (B) rules. The complete screening log is Table~\ref{tab:screening}.}
\label{tab:context}
\footnotesize
\begin{tabular}{@{}>{\raggedright\arraybackslash}p{4.2cm} >{\raggedright\arraybackslash}p{3.4cm} >{\raggedright\arraybackslash}p{5.9cm} >{\raggedright\arraybackslash}p{3.3cm}@{}}
\toprule
Pioneer(s) & Setting & Contribution & Reason not ranked\\
\midrule
J.~C. Maxwell; E.~J. Routh; A.~M. Lyapunov; A.~Hurwitz; N.~Minorsky & Universities, navy & Governor analysis; stability criteria; three-term control \citep{maxwell1868governors,routh1877treatise,lyapunov1992general,hurwitz1895bedingungen,minorsky1922directional} & P: before 1926\\
H.~S. Black; H.~Nyquist & Bell Laboratories & Negative-feedback amplifier; Nyquist criterion \citep{black1934stabilized,nyquist1932regeneration} & A: industrial career\\
J.~G. Ziegler; N.~B. Nichols; W.~R. Evans & Industry & PID tuning rules; root locus \citep{ziegler1942optimum,evans1948graphical} & A: industrial career\\
S.~F. Schmidt & NASA Ames & Kalman filter for Apollo navigation \citep{mcgee1985discovery} & A: government laboratory\\
J.~Richalet & Industry (Adersa) & Model predictive heuristic control \citep{richalet1978model} & A: industrial career\\
M.~Fliess & CNRS & Differential flatness \citep{fliess1992systemes,fliess1995flatness} & A: no faculty title found\\
G.~C. Devol; J.~F. Engelberger & Unimation & First industrial robot \citep{devol1961programmed,engelberger1980robotics} & A: industrial career\\
V.~Scheinman & Stanford AI Lab staff; industry & Stanford Arm; PUMA & A: no faculty title found\\
C.~A. Rosen & SRI International & Shakey, the first mobile robot to reason about its actions \citep{nilsson2009quest} & A: industrial research\\
S.~Inaba; F.~Wang & FANUC; DJI & Numerical control and robots; consumer drones & A: industrial careers\\
P.~G. Backes; L.~H. Matthies & Jet Propulsion Laboratory & Planetary-rover operations and vision & A: government laboratory\\
G.~Giralt & LAAS-CNRS & Mobile-robot architectures & A: no faculty title found\\
R.~S. Sutton; A.~G. Barto; N.~J. Nilsson; M.~Minsky; J.~McCarthy; R.~Reddy & Universities & Reinforcement learning; AI; founding of the CMU Robotics Institute (Reddy) \citep{sutton2018reinforcement,nilsson2009quest} & B: artificial intelligence\\
B.~W. Parkinson & US Air Force; Stanford & Architecture of the Global Positioning System & B: satellite navigation\\
\bottomrule
\end{tabular}
\end{table*}

\section{Corrected attributions}\label{app:corrections}
\setcounter{table}{0}

Several attributions that circulate in popular accounts, and that appeared in the authors' initial seed lists, did not survive verification. They are summarised in Table~\ref{tab:corrections} as a service to teachers and writers.

\begin{table*}[!t]
\centering
\caption{Common attribution errors and their corrections.}
\label{tab:corrections}
\footnotesize
\begin{tabular}{@{}>{\raggedright\arraybackslash}p{5.0cm} >{\raggedright\arraybackslash}p{11.4cm}@{}}
\toprule
Common claim & Verified account\\
\midrule
Sliding-mode control was introduced by Slotine & Variable-structure control and sliding modes were developed in the 1960s-1970s by Emelyanov, Utkin and co-workers, who received the 1972 Lenin Prize \citep{emelyanov1967variable,utkin1977variable,ipu2018emelyanov}. Slotine's contributions were sliding and adaptive control of manipulators and contraction analysis \citep{slotine1983tracking,slotine1987adaptive,lohmiller1998contraction}.\\
The zero-moment point is due to Kato & The ZMP concept is due to Vukobratović and Stepanenko \citep{vukobratovic1972stability,vukobratovic2004zero}. Kato's achievements were the WABOT-1 humanoid and the WL biped series \citep{kato1974information,takanishi1985realization}.\\
Lossless convexification lands Falcon~9 & G-FOLD, based on lossless convexification, was flight-tested on the Masten Xombie vehicle \citep{acikmese2007convex,scharf2017implementation}. SpaceX reports using CVXGEN-generated code \citep{blackmore2016autonomous}.\\
The Kalman filter ``enabled'' the Apollo landings & The filter was adapted for Apollo circumlunar navigation by Schmidt's group at NASA Ames \citep{mcgee1985discovery}. The guidance computer and GNC system were developed by Draper's MIT Instrumentation Laboratory \citep{hall1996journey,mindell2008digital}.\\
SLAM was invented by a single person & Stochastic mapping originated with Smith, Self and Cheeseman \citep{smith1986representation,smith1990estimating}. Localisation and SLAM were developed by Durrant-Whyte, Leonard and others \citep{leonard1991mobile,durrantwhyte2006simultaneous}.\\
Probabilistic roadmaps are Latombe's method & PRMs were introduced by Kavraki, Švestka, Latombe and Overmars, from Kavraki's doctoral work \citep{kavraki1996probabilistic}.\\
The Roomba is Brooks's invention & Brooks co-founded iRobot. The Roomba (2002) was developed by an iRobot team \citep{jones2006robots}.\\
W.~M. Wonham and K.~J. Åström are living emeriti & Wonham died on 14 May 2023 and Åström on 12 November 2026 \citep{wonham2023memorial,lund2026astrom}.\\
J.~J. Craig is a professor of robotics & No sustained university professorship was found, so Craig, author of a widely used textbook, is not ranked under our eligibility rule.\\
\bottomrule
\end{tabular}
\end{table*}

\bibliographystyle{elsarticle-harv}
\bibliography{references}

\end{document}